\documentstyle[11pt,aaspptwo]{article}

\lefthead{Fields}
\righthead{On the Evolution of the Light Elements}

\def\beq{\begin{equation}}
\def\eeq{\end{equation}}
\def\beqar{\begin{eqnarray}}
\def\eeqar{\end{eqnarray}}
\def\la{\mathrel{\mathpalette\fun <}}
\def\ga{\mathrel{\mathpalette\fun >}}
\def\fun#1#2{\lower3.6pt\vbox{\baselineskip0pt\lineskip.9pt
  \ialign{$\mathsurround=0pt#1\hfil##\hfil$\crcr#2\crcr\sim\crcr}}}

\def\he#1{\hbox{${}^{#1}$He}}
\def\li#1{\hbox{${}^{#1}$Li}}
\def\b1#1{\hbox{${}^{1#1}$B}}
\def\c1#1{\hbox{${}^{1#1}$C}}
\def\n1#1{\hbox{${}^{1#1}$N}}
\def\o1#1{\hbox{${}^{1#1}$O}}
\def\fe{\hbox{${}^{56}$Fe}}

\def\yp{\hbox{$Y_{\rm p}$}}
\def\hii{\hbox{\ion{H}{2}}}
\def\msol{\hbox{$M_\odot$}}
\def\sg{\hbox{$\sigma_{\rm g}$}}
\def\stot{\hbox{$\sigma_{\rm tot}$}}
\def\gav{\hbox{$\langle g_{3} \rangle$}}

\slugcomment{to appear in {\it The Astrophysical Journal}}

\begin{document}

\title{ON THE EVOLUTION OF THE LIGHT ELEMENTS \\
I.{} D, $^{\bf 3}$He, AND $^{\bf 4}$He}

\author{Brian D. Fields}

\affil{ Department of Physics \\
The University of Chicago, Chicago, IL  60637-1433 \\
\medskip
and \\
\medskip
Department of Physics \\
The University of Notre Dame \\
Notre Dame, IN  46556}

\setcounter{footnote}{0}

\begin{abstract}

The light elements D, \he3, \he4, and \li7 are produced in big bang
nucleosynthesis and undergo changes in their abundances due to
galactic processing.  Since one may observe most of these elements
only in contemporary environments, knowledge of the intervening
evolution is necessary for determining the observational constraints
on primordial nucleosynthesis.  Chemical and stellar evolution model
dependences in light element evolution are systematically investigated
via a comparison of 1460 possible chemical evolution scenarios and of
stellar nucleosynthesis yields, all of which have been selected to fit
solar neighborhood C, N, O, and Fe abundances as well as the observed
local gas density and gas mass fraction.  The light element evolution
and solar system yields in these models are found to span a wide
range, explicitly demonstrating the model dependence.  The range of
model dependence for D, \he3, and \he4 solar abundances is calculated,
and its sensitivity to the heavy element constraints is noted.  The
chemical evolution contribution to the uncertainty in the observed
primordial light element abundances is estimated, and the effects of
this uncertainty on big bang nucleosynthesis results are discussed.
The predictions for the light elements are found to be correlated; the
extent and physical origin of these correlations is discussed.  D and
\he3 evolution is found to have significant model dependence, however,
the dominant factor determining their solar and interstellar
abundances is their primordial abundance.  In addition, \he3 is found
to be very sensitive to the details of processing in low mass stars.
\he4 yields are shown to be very model dependent; in particular, both
the introduction of mass loss and the possibly very high \he4 stellar
yields in the poorly understood mass range of $\sim 8-12$ \msol\ can
lead to large enhancement of \he4 production and can lead to large
slopes of $\Delta Y/\Delta {\rm N}$ and $\Delta Y/\Delta {\rm O}$.  It
is found that the inclusion of secondary nitrogen leads to only a
small distortion in the low metallicity $Y -$ N relation if there is
also a significant contribution of primary N, as required by
observations.

\end{abstract}


\section{Introduction}
\label{sec:intro}

A central element of the hot big bang cosmological framework is the
prediction of light element (D, \he3, \he4, and \li7) synthesis during
the epoch of big bang nucleosynthesis (hereafter BBN).  This
prediction is tested against observations of the light elements
abundances, and the agreement between theory and observation provides
major empirical support for the big bang.  The comparison between
prediction and measurement is not immediate, however.  Many of the
observations are of environments that are contemporary to the present,
or nearly so, and galactic astrophysical production and destruction of
the light elements may have changed their abundances.  An
understanding of these galactic processes is thus needed to in order
to accurately interpret the light element observations and so to
realistically constrain BBN.  The study of the nuclear history of
galaxies is the domain of galactic chemical evolution; here we will
re-examine the galactic chemical evolution of the light elements.

Models for galactic chemical evolution were first constructed in the
1970's and for the most part studied the nucleosynthesis history of
the heavy elements.  Since such pioneering work as Cameron \& Truran
(\cite{camtru}), Talbot \& Arnett (\cite{ta}), and Tinsley
(\cite{tins,oldtins}), chemical evolution has come to do a
surprisingly good job of reproducing many solar abundance patterns and
even many abundance histories as observed in and around the solar
neighborhood (see, e.g., the recent work of Timmes, Woosley, \& Weaver
\cite{tww}; Matteucci \& Francois \cite{mfran}; and Ferrini et al.\
\cite{ferrini}).  These models also give reasonable predictions for
stellar luminosity evolution (see, e.g.~Tinlsey \cite{tins}; Colin \&
Schramm \cite{cs}), and underpin the study of nucleocosmochronology
(e.g.~Thielemann et al.\ \cite{cosmochron}).

The chemical evolution of the light elements has been studied since
the work of Cameron \& Truran (\cite{camtru}), and has been used to
determine primordial abundances since the work of Reeves, Audouze,
Fowler, \& Schramm (\cite{rafs}) and Audouze \& Tinsley (\cite{at}).
These seminal papers made a critical comparison of light element
abundances from deuterium to boron with the various cosmological and
astrophysical processes suggested for their production.  Since this
early work there have been many calculations of light element chemical
evolution, usually focusing on a few nuclides (see references in \S\S
\ref{sec:deu}--\ref{sec:he-4}).  In this paper we will again take a
comprehensive look at all of the light elements to see how consistent
a picture of their joint chemical evolution can be made.

Chemical evolution models are physically based on important mass
conservation principles, and have been refined considerably over the
years.  Unfortunately, though, chemical evolution is still a somewhat
ambiguous enterprise.  One should always be mindful of the
uncertainties in the conventional prescription for these calculations.
These uncertainties stem from unknown character of star formation
rates and mass distributions, galactic inflow and outflow of gas, and
{}from unresolved model dependent effects in stellar evolution
nucleosynthesis calculations.  Lacking a first principles calculation
of these phenomena, we are forced to model them in a phenomenological
or schematic way.

As a result of these uncertainties, the commonly adopted framework for
chemical evolution contains features that are not well understood in a
theoretical and/or empirical way.  Thus within this framework, there
are a great many ways to construct specific models, and work has been
done using a diverse variety of the possible scenarios.
Encouragingly, the overall framework has been quite successful despite
these difficulties.  And indeed chemical evolution is a always a part
of comparison of nucleosynthesis calculations with data (though
sometimes its role is only implicit).

The effects of chemical evolution model dependences have been
investigated to varying degrees.  However, the larger studies of these
effects (e.g.~that of Tosi \cite{tosi}) did not concentrate on the
light elements, and work on light element evolution has not examined
model uncertainties as broadly.  There has not yet been an emphasis on
the full range of uncertainty in a comprehensive description of light
element chemical evolution.

In this paper we construct a chemical evolution framework for the
evolution of all of lthe light elements, and we obtain estimates of
the range of model uncertainties.  We will follow the history of all
of the light elements, D through \b11.  We also include the
non-cosmological ``heavy'' elements \c12, \n14, \o16, and \fe, which
serve as tests of potential models.  We systematically examine a wide
variety of model assumptions, examining many chemical evolution
prescriptions and parameterizations (in the spirit of Tosi
\cite{tosi}), as well as including many suggested galactic production
and destruction mechanisms for each of the light elements (in the
spirit of Reeves, Audouze, Fowler, \& Schramm \cite{rafs}).

Our strategy is to choose a broad chemical evolution framework,
allowing for many features that authors have considered in examining
the histories of specific light elements.  These features are chosen
to be inclusive, without regard for their compatibility.  With these
we form a very large ($\sim$ 1200) set of models which include all
combinations of these features.  We then constrain these models to fit
observed solar neighborhood characteristics, apart from those of the
light elements.  This leaves $\sim 120$ potential models which are
viable descriptions of solar neighborhood chemical evolution (again,
light element behavior aside).  To each of these successful models we
add a range of 18 variants of light element evolution.  We then
examine the predictions of this large battery of light element
evolutionary schemes.  Specifically: (1) we see which models fit light
element observations; (2) we estimate the chemical evolution
uncertainty and discuss its influence on the use of BBN observational
constraints; and finally (3) we note correlations between the
evolution of different light element nuclides.  As part of this
analysis we will consider as well the uncertainties in the solar
system constraints themselves.

In addition to these considerations of the whole set of the light
elements, we will also address specific issues particular to each of
them.  For example, recent observations of high-redshift quasar
absorption line systems have approached the precision needed to give
meaningful D abundances.  This is an exciting prospect indeed, as
these high-$z$ environments are much more pristine than contemporary
ones, and so give a new window to a much earlier epoch in D evolution
(though still $\ga 1$ Gyr after BBN).  The initial reported abundances
by Songalia, Cowie, Hogan, \& Rugers (\cite{schr}) and by Carswell,
Rauch, Weymann, Cooke, \& Webb (\cite{crwcw}) were tantalizingly high
if taken at face value (see, e.g.~Kernan \& Krauss \cite{kk};
Cass\'{e} \& Vangioni-Flam \cite{cv}; Steigman \cite{steig}).
However, the observers themselves have been careful to caution that
the absorption line systems may be subject to contamination, and so
the derived D abundances are best understood as upper limits to the
actual levels of D/H, a caution strengthed by recent reports of a low
D abundance in a different line of sight (Tytler \& Fan \cite{tf}).
Clearly, the present situation is in flux and more observations are
needed to determine the D abundance in these systems accurately; but
just as clearly, this technique is quite promising and holds the key
for a very accurate D abundances at very early epochs.  We will
examine D evolution in detail in this paper, and comment on the role
of it chemical evolution in interpreting the quasar absorption line
system observational program.

These new constraints on D evolution also call attention to
conventional accounts of the evolution of \he3.  Recent improved
observations of \he3 in Galactic \hii\ regions can be puzzling in
light of preceding discussions of this element.  The data suggest that
\he3 shows spatial variations, and that the abundance of \he3 today in
the ISM is not significantly larger than that at the formation of the
solar system---indeed, in some places, \he3 is even seen in (possibly)
smaller amounts.  This may call into question (Hogan \cite{hogan}) the
conventional wisdom about \he3 production in low to intermediate mass
stars (Iben \& Truran \cite{it}).  On the other hand, the high \he3
abundance seen in planetary nebulae (Rood, Bania, \& Wilson
\cite{rbw92}) seems to support the notion of production in low mass
stars at a level roughly consistent with that predicted by Iben \&
Truran \cite{it}.  Thus the \hii\ region data may more likely point to
a need to re-examine \he3 evolution in high mass stars (Olive, Rood,
Schramm, Truran, \& Vangioni-Flam \cite{orstv}).  At any rate, the
issue of low mass stellar production of \he3 is crucial and we will
address it explicitly.

The much more abundant helium isotope, \he4, is important because it
is so powerful.  Notably, an accurate assay of the \he4 content of the
universe provides a bound on $N_\nu$, the number of light neutrino
families, as first shown by Steigman, Schramm, and Gunn (\cite{ssg}).
Thus there are continual improvements in the accuracy of theoretical
predictions for \he4 (e.g.~Dicus et al.\ \cite{dicus}; Walker et al.\
\cite{wssok}; Kernan \cite{kernan}; Seckel \cite{seckel}; Fields,
Dodelson, \& Turner \cite{fdt}), as well as in the observed abundances
(Skillman et al.\ \cite{skiletal}).  A pressing issue now is the
systematic uncertainty in these measurements (e.g.\ Copi, Schramm, \&
Turner \cite{cst}; Sasselov \& Goldwirth \cite{sg}). There is also
discussion of the correct means of extrapolating the primordial \he4
abundance, an issue in which chemical evolution plays a crucial role.

In this paper we will describe our framework for constructing models
of light element chemical evolution, and summarize our overall results
on model dependences.  We will then discuss in more detail our results
for D, \he3, and \he4 evolution and its uncertainties.  In a
forthcoming work (Fields \cite{fields}, hereafter Paper II) we will
discuss our results for Li, Be, and B evolution, with particular focus
on the range of available models for cosmic ray nucleosynthesis of
these elements.

\section{Light Element Evolution}
\subsection{BBN Predictions}
\label{sec:bbn}

The BBN predictions are crucial to our considerations, as the
primordial yields set the initial conditions for the chemical
evolution calculation.  The basic BBN calculation itself has not
changed dramatically since the calculation of Wagoner, Fowler, \&
Hoyle (\cite{wfh}).  For reviews we refer the reader to, e.g., Schramm
\& Wagoner (\cite{schwag}), Boesgaard \& Steigman (\cite{bsteig}), and
to Smith, Kawano, \& Malaney (\cite{skm}).  For recent calculations
see Walker et al. (\cite{wssok}), Krauss \& Kernan (\cite{kk}), and
Copi, Schramm \& Turner (\cite{cst}).

We consider only the simplest and ``standard'' case, namely that of
BBN in a universe that is spatially homogeneous in all particle
species.  Some models with baryon inhomogeneities are still allowed by
the data, but even these more complicated cases lead to the same basic
conclusions, as discussed in, e.g., Malaney \& Mathews (\cite{mm}) and
Thomas et al.\ (\cite{tsommf}).

In the standard calculation, the only free parameter is $\eta \equiv
n_B/n_\gamma$, the baryon-to-photon ratio.  In Figure \ref{fig:schplt}
we plot the light element abundances as a function of $\eta$; for
reference, Copi, Schramm, \& Turner (\cite{cst}) found concordance
between theory and data for the range \beq
\label{eq:etalim}
2.4 \times 10^{-10} \le \eta \le 6 \times 10^{-10} \eeq which we have
indicated on the plot.

For our purposes there are several salient features to notice.  First,
the \he4 abundance does not depend strongly on $\eta$, with variations
only at the 10\% level; however, we will be interested in the \he4
abundance to this level.  Even more evident is the decline in D and
\he3 abundance with increasing $\eta$, allowing a large range of
initial values, particularly for D which varies more strongly with
$\eta$.  This point is important, as we will find that the a
requirement of significant depletion of initially high D down to the
solar and interstellar abundance is a serious constraint on chemical
evolution models.

As we will see, the combined abundance D+\he3 is a most useful
constraint on $\eta$.  Note that for the concordant range of $\eta$
that this sum is dominated by the contribution from D.  Knowledge of
the primordial level of D is tantamount to knowledge of D+\he3; thus
the D abundances obtained from quasar absorption line systems provides
a constraint on this sum.  Thus an accurate quasar measurement would
provide a crucial complement to the solar system and interstellar D
and \he3 abundances.

\subsection{Chemical Evolution Framework}

The basic theoretical framework and observational constraints on
models of chemical evolution is reviewed in Tinsley (\cite{tins}).  We
will, for the most part, adopt her notation.  Our goal is to trace the
history of the Galaxy's isotopic content.  Ideally, we wish to model
formation of stars in inhomogeneously distributed gas clouds, their
return of material during their lifetime to the ISM via winds, and
finally their deaths, which occur in stochastic bursts and return
processed and unprocessed material to the ISM.  In principle, all of
these processes can depend on many variables, e.g.~the metallicity of
the gas, the temperature of the gas, and the strength and
configuration of the galactic magnetic fields, to name a few.

However, this depiction is not computationally feasible.  To reduce to
problem to a tractable one, we will employ several simplifying
assumptions: (1) the Galaxy (or at least the solar neighborhood) is
spatially homogeneous; (2) stars are born according to a given mass
distribution, and their lifetimes are a function of mass only; and (3)
all of the stellar ejecta are returned at once at the time of the
star's death, and instantaneously mixed in the interstellar medium
(ISM).  These assumptions, though strong, are adequate for many
applications, and in any case are necessary to proceed.

We will employ the standard convention for the stellar creation rate
$C$, which gives the number of stars born (per pc${}^2$) per Gyr, in
dividing it into a star formation rate (hereafter SFR, $\psi$) and an
initial mass function (hereafter IMF, $\phi$)
\beq
\label{eq:cfun}
C(m,t) \, dm \; dt = \psi(t) \phi(m) \, dm \; dt \; \; .
\eeq
The SFR
has units of [$\msol {\rm Gyr}^{-1} {\rm pc}^{-1}$], and gives the
mass of ISM material going into stars per unit time.  The IMF gives
the mass distribution with which these stars are formed.  Questions of
the time constancy of IMF have been to answer theoretically or
observationally (see, e.g.~Scalo \cite{scalo}), although recent work
(Beers, Preston, \& Shectman \cite{bps}) supports the conventional
assumption of its constancy, which we will follow.

The literature is fraught with
unfortunate variations in the convention for IMF normalization;
we follow Tinsley (\cite{tins}) in choosing
\beq
\int_{m_{\rm low}}^{m_{\rm up}} \; dm \, m \; \phi(m) \equiv 1
\eeq
The
lower and upper mass limits, $m_{\rm low}$
and $m_{\rm up}$, are not well constrained by
data, and so will be parameters we will have to choose.

We will use a one zone model for the solar neighborhood,
treating it as homogeneous and
instantaneously
mixed.   We will track the total disk mass in terms of its
surface mass density $\sigma_{\rm tot}$, which evolves by
\beq
\frac{d\sigma_{\rm tot}}{dt} = f
\eeq
where $f$ is the rate for the infall of new material due to, e.g.~the
collapse of the disk or to accretion of extragalactic material, in units
[\msol Gyr$^{-1}$ pc$^{-2}$].

We put the basic constituents of matter as stars (including their
remnants) and gas,
\beq
\sigma_{\rm tot} = \sigma_{\rm s} + \sg \; \; .
\eeq
Of these, we will explicitly track the gas, which evolves according to
\beq
\frac{d\sg}{dt} = - \psi + E + f
\eeq
where
\beq
E(t) = \int_{m_t}^{m_{\rm up}} \; dm \; m^{\rm ej} \; \phi(m) \;
\psi\left(t-\tau_m\right)
\eeq
is the rate at which dying stars eject processed material back into
the ISM.  The lower mass limit, $m_t$, is the stellar mass $m$ for
which $\tau_m = t$, i.e.~the smallest mass star which dies just at
time $t$ after being created at $t=0$.

We will define the mass fraction
\beqar
X_i \equiv \frac{\sigma_i}{\sigma_{\rm tot}} \; , & \hbox{\hskip 2em}
& \sum_i X_i \equiv 1
\eeqar
and we write
\beq
\label{eq:dxidt}
\frac{d \left(\sg X_i\right)}{dt} = - \psi X_i + E_i
+ f X^{\rm inf}_{i}
\eeq
where
\beq
E_i(t) = \int_{m_t}^{m_{\rm up}} \; dm \; m_i^{\rm ej} \phi(m)
\psi\left(t-\tau_m\right)
\eeq
is the rate of the ejection of isotope $i$ to the ISM.

Note that the presence of the ``retarded time'' $t - \tau_m$ makes the
equations unwieldy.  In practice one often makes $\psi$ a function of
$\sg$ and/or $\sigma_{\rm tot}$, thus rendering the basic equations as
integro-differential and hence without an analytic solution We have
therefore numerically implemented the preceding equations, taking into
account their full integro-differential nature.  Our results are all
based on these calculations.

For comparison with calculations using
the instantaneous recycling approximation
(IRA), it is useful to define the return fraction
\beq
R = \int_{m_{\rm low}}^{m_{\rm up}} \; dm \; m^{\rm ej} \phi(m)
\eeq
the fraction of mass going into a generation of stars that will
eventually be ejected back into the ISM.  Note that $R < 1$ and that
it is very sensitive to the choice of IMF form and mass limits.  We
will also use $\mu \equiv \sg/\sigma_{\rm tot}$ for the gas mass
fraction.

Of the model features we have described, those most influential on the
abundance predictions are the IMF, the SFR, infall, and the adopted
nucleosynthesis yields.  Of these, the IMF, SFR, and infall (or
outflow) are not well understood theoretically, and so one must select
a phenomenological prescription for each.  The stellar nucleosynthesis
inputs come from model calculations for different stellar masses (and
sometimes different metallicities).  For the most part only a limited
number of metallicities are used, thus forcing chemical evolution
models to interpolate and extrapolate.  A better treatment would
iteratively run both stellar models and chemical evolution models;
such a program has recently been done by Timmes, Woosley, \& Weaver
(\cite{tww}), who emphasized heavy element yields which they
calculated self-consistently using a supernova code.

Of the other input parameters, two are cosmological.  The first is of
course the BBN result of the initial abundances; our approach to this
is discussed below (\S \ref{sec:result}).  The other cosmological
parameter is the age of Galaxy including the collapse of the disk; we
will take this 15 Gyr, and our results are not very sensitive to this
choice.  We will take the age of the earth to be 4.6 Gyr.  Finally, we
will use the $\tau_m$ vs.~$m$ relation of Scalo (\cite{scalo}).

Having selected a set of models representing the range of choices of
the major model features---IMF, SFR, and nucleosynthesis---we will
want to constrain this set with observational data.  Specifically, we
will test these against solar system abundances of CNOFe, as well as
the gas mass fraction.  The confrontation with observations will
select a set of viable models which we will then use to test different
models of light element abundances.  Thus we will want to include in
our initial suite of models all variations of the major features---
the IMF, SFR, and nucleosynthesis---which have been proposed to be
well-suited to fit individual light elements.  Therefore, we now will
review the chemical evolution of the light elements, to identify
features to include in the initial set of chemical evolution features.

\subsection{Deuterium}
\label{sec:deu}

\subsubsection{D Data}

Traditionally, D has been measured in the solar system and in the ISM.
Important evidence for the solar system D abundance came as a result
of the Apollo 11 mission (Geiss \& Reeves \cite{geiree}), giving a
measurement of the D+\he3 content in the solar wind.  These data are
complemented by measurements of meteorites (Black \cite{black}), which
allows determination of the presolar D level.  The best numbers now
are that the presolar D abundance is (D/H)$_\odot = 2.6 \pm 1.0$
(Geiss \cite{geiss}).

The D abundance in the ISM was first measured by Rogerson \& York
(\cite{ry}).  It has most recently been observed by the {\it Hubble
Space Telescope}, with Linsky et al.\ (\cite{linsky}) reporting
(D/H)$_{\rm ISM} = 1.5^{+0.07}_{-0.18}$.  This single, remarkably
precise measurement will prove to set strong constraints on chemical
evolution models.

Most D measurements to date have been of local and relatively recent
cosmological epochs.  However, exciting new observations have reported
possible detection of D in quasar absorption line systems (QSOALS).
These systems considered lie at high redshift ($z > 3$), and so are
very young (1--3 Gyr).  Thus one expects D to have suffered much less
depletion.  Indeed, initial reports (Songalia, Cowie, Hogan, \& Rugers
\cite{schr}; Carswell, Rauch, Weymann, Cooke, \& Webb \cite{crwcw})
suggested the possibility of a very high D, at levels of D/H $\la 2.5
\times 10^{-4}$.  Both of these groups have been careful to caution
that this result can only be regarded as an upper bound on D, as low
column density in intervening clouds can contribute significantly to
the putative D line.  Indeed, this may be the case for these
measurements, as Tytler \& Fan (\cite{tf}) have recently observed a
different line of sight and in a high-redshift absorption system claim
a preliminary D abundance of D/H $\simeq 2 \times 10^{-5}$.  The
situation may be more muddled still, as it has recently been suggested
(Levshakov \& Takahara \cite{lt}) that improper modeling of the
turbulent characteristics of these systems could lead to very large
erros in the derived D abundance.  We will thus take the prudent route
of not using the few D abundances obtained in this manner, and instead
concentrate on better-determined the solar and interstellar values.

\subsubsection{D Chemical Evolution}

Deuterium was originally thought not to be primordial, but to be
produced in the T-Tauri phase of stellar evolution (Fowler,
Greenstein, \& Hoyle \cite{fgh}).  However, this process was shown to
fail (Ryter, Reeves, Gradstajn, \& Audouze \cite{rrga}), and Reeves,
Audouze, Fowler, \& Schramm (\cite{rafs}) argued that D was likely of
cosmological origin.  This argument was cemented by the work of
Epstein, Lattimer, \& Schramm (\cite{els}), who showed that not only
does stellar burning destroy D, but {\it any} high energy
astrophysical process destroys it as well---except the big bang.

Indeed, as recently emphasized by Copi, Schramm, \& Turner
(\cite{cst}), D is best baryometer of BBN in that the constraints on
$\eta$ which arise from D observations are particularly reliable.
This reliability stems from the uniquely straightforward evolution of
D in galaxies: all of the D in the universe was produced in BBN, and
all subsequent processes destroy it.  This fundamental fact greatly
simplifies the chemical evolution of D, which has no sources and so is
absent from any processed material that is returned to the ISM.  The D
abundance thus provides a clean indication of the amount of material
that has not been processed in stars.  (The galactic evolution of D
\he3, and \he4 is summarized in table \ref{tab:source}).

To get a feel for D evolution, we note that in the instantaneous
recycling approximation, the D evolution can be solved exactly as
\beq
\label{eq:dev}
\frac{X_{2}}{X_{2}^{\rm p}} = \mu^{-R/(1-R)}  \; \; .
\eeq
Note that this result does not depend on the form of the star
formation rate $\psi$.  Furthermore, since the the gas fraction $\mu$
is constrained by observation, the entire character of the depletion
comes from $R$, the return fraction.  This is in turn completely
determined by the initial mass function.  Physically, the D abundance
is a measure of the fraction of gas which remains unprocessed.  Thus
as material is cycled into stars, the D depletion will depend on how
much (D-free) material is returned to the ISM to dilute the
unprocessed component.

The main chemical evolution feature to consider in order to capture
the range of D evolution is that which determines the return fraction
$R$, namely the IMF.  We will want to consider the gamut of possible
IMF forms and mass limits.  Indeed, we will expect all the light
element evolution to depend strongly on the adopted IMF.

\subsection{Helium--3}
\label{sec:he3}

\subsubsection{\he3 Data}

The solar system information on both \he3 and D come together and are
derived from measurements of the solar wind and meteorites (Geiss
\cite{geiss}; Black \cite{black}); the presolar abundance is
(\he3/H)$_\odot = 1.5 \pm 0.3$.  In the ISM, \he3 is observed in
Galactic \hii\ regions, and measured via its hyperfine line.  Although
these measurements are difficult, they apparently show real
variations, with (\he3/H)$_{\rm ISM} \sim (1-5) \times 10^{-5}$
(Balser, et al.\ \cite{balseretal}).  \he3 has also been detected by
Rood, Bania, \& Wilson (\cite{rbw92}) in a planetary nebula, and is
found at very high levels: $\he3/{\rm H} \sim 10^{-3}$.

\subsubsection{\he3 Chemical Evolution}

As noted in \S \ref{sec:he3}, BBN yields a fairly low \he3 abundance,
but not much lower than contemporary observations; i.e.~BBN theory
does not leave a lot of room for the increase of \he3.  On the other
hand, stellar models firmly show that all initial D in a star is
processed to \he3 in the pre-main sequence phase when the star is
fully convective.  Some of this \he3 survives to be ejected at the
star's death, and indeed in low mass stars are likely to be \he3
sources.  Thus while D suffers astration and \he3 is produced over
time, the sum of the two, D+\he3, is more stabile; furthermore, D+\he3
is first dominated by D and then \he3 increases until roughly equal
measures of each comprise the present ISM abundance.

Yang et al.\ (\cite{ytsso}) get a bound on (D+\he3)$_{\rm p}$ by
noting that in all stars, any initial D goes to \he3 in the pre-main
sequence phase.  Some fraction $g_3(m) < 1$ of that \he3 survives in
the ejecta at the star's death.  If we consider one stellar
generation, and let \gav\ be the average of $g_3(m)$ over the mass
function, it may be shown that
\beq
\label{eq:yang}
\left(\frac{{\rm D}+\he3}{\rm H}\right)_{\rm p}
\le \frac{1}{\gav} \left(\frac{\he3}{\rm H}\right)_{\rm 1} +
\left(\frac{\rm D}{\rm H}\right)_{\rm 1}
\eeq
(Yang et al \cite{ytsso}).  The inequality (eq.~\ref{eq:yang}) comes
from disregarding \he3 production, and so guaranteeing that $X_{23}$
decreases with time.

We may get another perspective on this point by seeing how the D+\he3
argument plays out not for a single stellar generation, but in the
approximation of instantaneous recycling (see preceding section) for a
closed box chemical evolution model.  We have
\beq
\frac{d \left(\sg X_3 \right)}{dt} = - \psi X_3
+ R \gav \psi \left( \frac32 X_{\rm D} + X_3 \right)
\eeq
and when substituting from eq.~\ref{eq:dxidt}
we have
\beq
\nonumber
\frac{(\frac32 X_2 + X_3)}{(\frac32 X_2 + X_3)_{\rm p}} =
\left(\frac{X_2}{X_2^{\rm p}}\right)^{1-\gav}
\eeq
and so in term of observable number ratios,
\beqar
\nonumber
\hphantom{\hskip 1.5em}
\frac{\left( {\rm D} +\he3 \right)}
     {\left({\rm D}+\he3 \right)_{\rm p}}
   & = &  \left(\frac{{\rm D}}{{\rm D}_{\rm p}}\right)^{1-\gav} \;
\left(\frac{X_1}{X_{\rm p}}\right)^{\gav} \\
& \approx &
\left(\frac{{\rm D}}{{\rm D}_{\rm p}}\right)^{1-\gav}
\eeqar
D+\he3 increases (decreases) with time if $\gav > 1$ ($\gav < 1$).
Note, however, that because low mass stars are crucial here, that the
IRA is inappropriate.  At any rate, there is a close relationship
between D and \he3, and we see that it is crucial to know whether \he3
is created or destroyed with time.

The physical explanation for the different fate of \he3 in high and
low mass stars is that for $m < 2 \msol$, $p-p$ is the dominant form
of hydrogen burning, and the $p(p,\nu e^+)$D($p,\gamma$)\he3 chain
makes \he3.  However, for $m > 2 \msol$, the CNO cycle dominates the
hydrogen burning, and \he3 is destroyed.  Indeed, Iben \& Truran's
(\cite{it}) calculation for low mass stars suggests that \he3 is
produced copiously.  Dearborn, Schramm, \& Steigman (\cite{dss}) find
that \he3 is destroyed in high mass stars, particularly at low
metallicity.  Woosley \& Weaver (\cite{wwprep}) give more detailed but
qualitatively similar results.

These conventional notions about \he3 evolution have recently been
called into question as the interstellar \he3 abundances have become
more accurate.  Until recently, the production of \he3 in low mass
stars was not considered in chemical evolution studies, e.g.\ that of
Steigman \& Tosi (\cite{st}).  Vangioni-Flam, Olive, and Prantzos
(\cite{vop}) let the contribution to $\gav$ from low mass stars be a
free parameter, and used the Dearborn et al.\ (\cite{dss}) high mass
yields.  They found that to fit the data require net destruction by
low mass stars, a mechanism for which is suggested by Hogan
(\cite{hogan}).  Indeed, with the Iben \& Truran (\cite{it}) and
Dearborn et al.\ (\cite{dss}) results, Galactic \he3 which starts with
the espoused BBN value is apparently overproduced in many chemical
evolution models (Olive et al \cite{orstv}; Galli et al.\
\cite{gpsf}).  However, in the face of this apparent need for \he3
destruction is the observed prescence of high \he3 in planetary
nebulae (Rood, Bania, \& Wilson \cite{rbw92}) at a level in rough
agreement with Iben \& Truran's calculation.

Another \he3 puzzle is the observed and apparently real dispersion in
\he3 abundances.  Balser et al.\ (\cite{balseretal}) present this
dispersion as a \he3 abundance gradient, with \he3 decreasing towards
the galactic center.  This trend runs counter to intuition, if \he3 is
to be formed by low mass stars which are more prevalent towards the
galactic center.  Moreover, for stars having lifetimes this long, it
is hard to see how this gradient could happen at all, as such stars
will have traveled a significant fraction of disk in their lifetimes
and so should be well spread throughout the Galaxy.

A possible explanation for ``what is wrong with \he3'' was offered by
Olive et al.\ (\cite{orstv}).  They point out that the correlation of
\he3 with galactocentric distance might be better understood as a
correlation with the mass of the \hii\ region.  Then the \he3 in the
higher mass regions will contain more ejecta from high mass stars.  In
this scenario, one expects \he3 to be lower wherever there is more
ejecta from massive stars, i.e. more stellar processing.  Thus the
\he3 abundance should go down towards the center of the galaxy, where
the density and so the star formation rates are higher.  Note that
this could also answer the puzzle of the existence of the gradient
despite the long lifetime of low mass stars.  The point is that there
is not a gradient in \he3 production (by low mass stars), but instead
there is a gradient in destruction by short--lived, high mass stars.

For our chemical evolution models, we will want to use our apparatus
to address the important issue of the effect of low mass processing on
\he3 evolution, allowing for different possibilities as done by
Vangioni-Flam, Olive, \& Prantzos (\cite{vop}).  We will also include
infall, as was done in Steigman \& Tosi (\cite{st}).

\subsection{Helium--4}
\label{sec:he4}

\subsubsection{\he4 Data}
\label{sec:he-4}

Observations of {}\he4 are well reviewed elsewhere, e.g.~Pagel
(\cite{pagel}), Pagel et al.\ (\cite{pageletal}), Walker et al.\
(\cite{wssok}), Davidson \& Kinman (\cite{dk}), Skillman \& Kennicutt
(\cite{skilken}); Skillman et al.\ (\cite{skiletal}); and Copi et al.\
(\cite{cst}).  Briefly, accurate determinations of \he4 are hard to
come by.  The best observations are of \hii\ regions, where He should
be ionized and accurate abundances might result.  Although there have
been observations of \he4 in Galactic \hii\ regions, systems with
lower metallicity are expected to contain less contamination from
stellar ejecta.  The sites of choice have proven to be metal poor
extragalactic \hii\ regions.\footnote{Jakobsen et al.\ (\cite{jak})
have recently reported detection of \he4 in a quasar absorption line
system.  The uncertainties in the measurement render it a qualitative
confirmation of BBN, but with more accuracy this type of observation
could be extremely powerful.}

The solar system abundance of \he4 is $Y_\odot = 0.274 \pm 0.016$
(Anders \& Grevesse \cite{ag}).  ISM abundance determinations are few,
and furthermore, the ionization structure of gas regions causes \he4
to be systematically underestimated.  (Wilson \& Rood \cite{wr}).
Thus, we have no information on \he4 in the ISM that is accurate
enough to be an important constraint on chemical evolution models.

\subsubsection{\he4 Chemical Evolution}

As the second most abundant nuclide in the universe, next to hydrogen,
\he4 comprises about a quarter of the universe's baryonic mass.  It is
also the only nuclide made in significant amounts in stars of a broad
range of masses, with the dominant production in stars of middle
masses (Renzini \& Voli \cite{rv81}; Iben \& Truran \cite{it}).
Despite this nearly ubiquitous stellar synthesis, however, stars only
make a small ($< 10 \%$) contribution to the total helium abundance
(Hoyle and Tayler \cite{ht}): most \he4 is primordial.  Nevertheless,
to fully test BBN and moreover to use its power to constrain particle
physics and/or test conditions in the early universe, one needs to
know the abundance of \he4 exceedingly well.

Even in low metallicity \hii\ regions, stellar pollution exists; the
question is how best to determine it and so deduce the primordial He
abundance.  To this end Peimbert \& Torres-Peimbert (\cite{ptpa,ptpb})
noted that He production should be correlated with heavier element
production.  In \hii\ regions one observes both He and CNO.  Following
Peimbert \& Torres-Peimbert, one deduces the primordial helium mass
fraction \yp\ from extrapolating the low metallicity end of the $Y$
vs.\ $Z$ plot, exploiting the relation
\beq
\label{eq:dydz}
Y = \yp + \frac{dY}{dZ} \; Z
\eeq
Equation (\ref{eq:dydz}) should be valid for sufficiently small $Z$
(and $dY/dZ$ a constant function of $Z$), for which this procedure
should succeed in extrapolating \yp.  The challenge for chemical
evolution is to determine which $Z$ are sufficiently small, and which
``metals'' C, N, and/or O are the best surrogates for $Z$.  The
challenge for observers is to make precise enough measurements to
allow for a meaningful extrapolation.\footnote{The difficulties of
measuring He and metal abundances to the desired accuracy (\yp\ good
to the third decimal place!)  cannot be overstated.  For attempts to
correct for some of these errors by culling points from the data set
see Pagel (\cite{pagel}), Pagel \& Kazlauskas (\cite{pagkaz}), Pagel,
Simonson, Terlevich, \& Edmunds, (\cite{pageletal}), and Olive \&
Steigman (\cite{os}).}

Analysis using this technique has for the most part avoided the
explicit use of detailed chemical evolution calculations.  Instead,
the approach has been an empirical one: $Y$ and $Z$ are measured for
different regions; the $Y$ vs.~$Z$ relation is plotted, fit, and
extrapolated to get $\yp = Y(Z=0)$.  Two complications to this
procedure immediately arise.  First, one does not directly measure the
metal fraction $Z$, and so one uses as a surrogate either oxygen or
nitrogen (and carbon---see Steigman, Gallagher, \& Schramm
\cite{sgs}).  Second, and moreover, one must adopt an expected $Y$
vs.~O (and N or C) relation to use in fitting the data and making the
extrapolation.

While everyone recognizes that the fits need not be linear, this is
the simplest two parameter fit, and so this has been the first to be
tried.  A glance at the data (see figure \ref{fig:yno}) suggests that
this might not be a bad first guess.  Indeed, the $Y$ vs.~O/H relation
is well described by a linear fit.  In contrast, the case of nitrogen
as a metallicity tracer has generated vigorous debate as to the
appropriate fitting function.  At issue is whether to choose a linear
fit, in which the slope $\Delta Y/\Delta N$ is constant, or a
nonlinear one, in which ${\rm N} \propto Y^2$, and thus the slope is
not constant and has the possibility of being quite steep for small N,
i.e.\ at early times (Fuller, Boyd, \& Kalen \cite{fbk}; but see
Olive, Steigman, \& Walker \cite{osw}).

Put differently, the question regarding the $Y-{\rm N}$ relation is,
what is the N $-$ O relation?  The nonlinear fit for $Y$ vs.~N arises
by assuming that N production is proportional to the C and O
abundance, leading to the ``secondary'' relationship N $\propto {\rm
O}^2$.  We wish to know the relative size of the primary and secondary
stellar contributions to N, an issue not yet resolved.  Consequently,
we will want to test effects of both primary and secondary N.

Most \he4 observations are determined from extragalactic \hii\
regions.  To completely understand these abundances would require
separate evolutionary models for each observed galaxy.  However, we
will be fitting our model to the solar system, and so we should not
necessarily expect a close fit to the extragalactic data.  However,
the few models that do exist, e.g.~those of Mathews Boyd \& Fuller
(\cite{mbf}), and Balbes, Boyd, \& Mathews (\cite{bbm}), are
qualitatively similar to those studied here.

Another outstanding problem in \he4 evolution is the theoretical
reproduction of the slope $\Delta Y/\Delta Z$.  The theoretical
$dY/dZ$ depends on how many high mass stars are included in the
calculation (e.g.~Maeder \cite{mae83,mae92,mae93}; Brown \& Bethe
\cite {bb}; and Prantzos \cite{pran}).  Stellar yields for massive
stars give a much lower ratio $dY/dZ \sim 1.3$ than that for
intermediate and low mass stars, $dY/dZ \sim 6$.  Averaged over a
typical IMF with mass limits of 0.4 and 100 \msol\ gives
\beq
dY/dZ  \sim 2
\eeq
at most (Maeder \cite{mae92,mae93}).  This theoretical slope is to be
compared with an observed slope (calculated assuming $Z \propto$ O) of
$dY/dZ \sim 4-6$ (Pagel \cite{pagel}).

One attempt to address this issue notes that many models for high mass
stars (e.g.~Weaver \& Woosley \cite{ww93}) do not include mass loss.
For stars above $\sim 30$ \msol\ this can have a significant impact on
all abundances but particularly \he4, as suggested by the results of
Maeder (\cite{mae92}).  Indeed, stars above $\sim 30$ \msol\ are
thought to rapidly lose mass until reaching $30 \msol$; thus one need
only compute stellar yields up to this mass.  For our purposes, we
will explicitly include the possibility of mass loss in high mass
stars.

Finally, an unfortunate aspect of stellar evolution is that there is a
great deal of uncertainty as to the gross behavior--let alone the
nucleosynthesis yields--of stars in the 8--10 \msol\ mass range
(further discussed below, \S \ref{sec:mod-nuke}).  It has been
suggested (Woosley \& Weaver \cite{wwrev}) that these stars might
dominantly produce helium.  This could have a large effect on the net
\he4 production and so it is worth trying $8-10 \msol$ yields with
very large yields helium, to see the sensitivity to this range.

\section{The Suite of Models}

Our goal is to examine a broad selection of physically plausible
scenarios, including at least the variations that have been commonly
used in the literature.  This approach is inspired by that of Tosi
(\cite{tosi}), although her study made infall a central feature
whereas we will emphasize other features as well.

We classify model features as ``global'' in scope for particular to
the light elements.  We will designate ``global'' model features to
be: (1) those that affect all elements (e.g.~choices of SFR or IMF)
and the gas consumption; or (2) those that affect the yields of C, N,
O, and Fe (CNOFe), i.e.~the nuclides we will be following which are
not light elements but serve as chronometers or tracers for the light
elements.  Upon identifying the set of global model variations, we
combine all of these features independently to create a large set of
potential models for the solar neighborhood.  By systematically
investigating every possible combination of our chosen global
features, we will be assured to properly estimate the full model
variation inherent within the span of model options we have allowed.

We demand that our results conform to the features of the solar
neighborhood.  By selecting those models which reproduce the solar
characteristics, we will develop a suite of acceptable global models.
To each of these global models is then added a wide variety of light
element evolutionary prescriptions.  We then calculate the chemical
evolution for every light element variant, obtaining an inclusive
measure of the model dependence of light element evolution.

\subsection{Selection of Global Model Features}

In her study, Tosi (\cite{tosi}) found that all aspects of chemical
evolution models affect the absolute abundances.  However, she found
that the choice of IMF, and of course the selection of nucleosynthesis
yields, had the most profound influence, affecting the abundance
ratios as well as the absolute abundances.  Therefore we will want to
be particularly thorough in our consideration of these model features.
Our global model properties are summarized in table \ref{tab:gmodf}.

\subsubsection{Nucleosynthesis Yields}
\label{sec:mod-nuke}

The tabulated nucleosynthesis yields are those of low to intermediate
mass stars, with masses $\sim 0.8 \msol$ to $\sim 8 \msol\ $, and high
mass stars with masses $M > 12 \msol\ $.  The former are understood to
be stars whose lives progress to the point of planetary nebula
ejection and becoming carbon-oxygen white dwarfs; the latter, high
mass stars end their lives explosively as type II supernovae.
Standard results for this study are those of Renzini \& Voli
(\cite{rv81}) for intermediate mass stars, and Weaver \& Woosley
(\cite{ww93}) for high mass stars.

The uncertainties in the yields pointed out in these references.  The
Renzini \& Voli (\cite{rv81}) results, for example, were found to
depend on the adopted parameterization of convection in the form of
the mixing length.  However, this level of detail is not a major
source of uncertainty for our purposes.  It would be desirable to
compare results for different stellar models and yields; unfortunately
this is not possible for the low mass stars, as Renzini \& Voli
(\cite{rv81}) are the only current models with detailed results
reported.  In addition, for the massive stars, several groups have
published massive star yields; however these models have very
different degrees of emphasis on their nuclear yields, with Woosley \&
Weaver (\cite{ww93}) presenting particularly meticulous results.

As motivated in \S \ref{sec:he4}, we allow for a wide range of N
evolution, including both primary and secondary sources for N.  We
thus will follow the approach of Mathews, Boyd, \& Fuller (\cite{mbf})
in writing the mass fraction of ejected N, $X^{\rm ej}$, as
\beq
X^{\rm ej} = X_0^{\rm ej}\left(\alpha
    + \beta \frac{X_C(t)}{X_C^\odot}\right)
\eeq
where $X_0$ is the usual N yield, $X_C(t)$ is the calculated C mass
fraction and $X_C^\odot$ is the solar carbon mass fraction from Anders
\& Grevesse (\cite{ag}).  The constants $\alpha$ and $\beta$ are
chosen as follows:
\beq
(\alpha,\beta)  =  (1,0) \ ; \ (0.5,1) \ ; \
   (0,1) \ ; \ (0,2)
\eeq
where the first option is the standard case of purely primary N, the
second is a mixed case of some primary and some secondary N, and the
final two cases are variation on purely secondary N.  Note that our
procedure differs somewhat from that of Mathews, Boyd, \& Fuller
(\cite{mbf}) who used O as the seed nucleus for secondary N, instead
of C.  Carbon is the more appropriate choice (Audouze, Lequeux, \&
Vigroux \cite{alv}; Vigroux, Audouze, \& Lequeux \cite{val}; and
Dearborn, Tinsley, \& Schramm \cite{dts}), and we have used it.

\subsubsection{Initial Mass Function}

The selection of an initial mass function is a crucial feature of any
chemical evolution model.  Unfortunately, there is no convincing
physical theory of star formation, and so the nature of the IMF and
the star formation rate over the history of the Galaxy are poorly
understood.  It is unclear, for example, whether the IMF has changed
with time, what its upper and lower mass limits are (and have been),
and whether it depends upon the composition of the gas which will
become stars.  Furthermore, it is very difficult to untangle the
behavior of the IMF and the SFR, which could in fact be inseparable
and better left in the form of a creation function, as in
eq.~(\ref{eq:cfun}).

Lacking any good theoretical or observational guidance as to these
questions, we will adopt eq.~(\ref{eq:cfun}), i.e.~we will assume that
the IMF is constant in time.  We will, however, allow for it to take
different forms.  We will try the classic Salpeter (\cite{salp})
function,
\beq
\phi(m) \propto m^{-(1+x)}
\eeq
and we will allow the index (``slope'') $x$ to vary, trying
$x=1$,1.35, and 1.7.  We will also try an IMF $\phi$ derived from the
observed, present-day mass function (PDMF, $\phi^{\rm pd}$), as given
by, e.g.~Scalo (\cite{scalo}).  The relationship between the two
depends on whether the stellar lifetime $\tau_m$ is short enough that
some of the stars of mass $m$ have died; specifically,
\beq
\label{eq:pdmf}
\phi(m) = \left\{
	\begin{array}{ll}
	\phi^{\rm pd}(m) \, , & \tau_m \ge t_0 \\
	\phi^{\rm pd}(m)/\int_{t_0-\tau_m}^{t_0} dt \; b(t) \, ,
              & \tau_m < t_0
	\end{array}
\right.
\eeq
where the age of the Galaxy is $t_0$ and $b(t) = \psi(t)/\langle \psi
\rangle$.  We remind the reader that this procedure is only
practicable when the star formation rate $\psi$ is a given, explicit
function of time,\footnote{In fact, other SFRs can accommodate the
Scalo IMF through an iterative process (Mathews, Bazan, \& Cowan
\cite{mbc}); such a procedure would however be very time consuming to
use with our many models, and is not considered here.}  because of the
role of $\psi$ in eq.~(\ref{eq:pdmf}).

We also will try different mass limits to the IMF.
Here we are guided by common choices that have appeared in
the literature.  We will wish to have a broad, medium, and narrow
range in mass.  We choose limis of
\beq
(m_l,m_u) = (0.2,100); (0.1,60); (0.4,30)
\eeq
which we feel to be representative.

\subsubsection{Star Formation Rate}

As the physics of star formation is ill understood, the form of the
SFR is not well set by theory, although there have been some attempts
to do so.  As with the IMF, there are many variants, but we will
select what we feel to be typical choices:
\beqar
\psi =  & \nu  \sigma_{\rm tot} \left( \frac{\sg}{\stot} \right)^0
  & = \nu \sigma_{\rm tot} \\
\psi =  & \nu  \sigma_{\rm tot}\left(\frac{\sg}{\stot}\right)^1
  & = \nu \sigma_{\rm g} \\
\psi =  & \nu  \sigma_{\rm tot}\left(\frac{\sg}{\stot}\right)^2 & \\
\psi =  & a \ \exp ( - t/\tau ) &
\eeqar
where we pick the $\tau = 7.5$, 15 Gyr as the timescale for the
exponential case.  Note that these always decrease for a closed box
($\sigma_{\rm tot} =$ const), but that for infall models with
accreting gas, these will show a rise at early times.

\subsubsection{Infall}
\label{sec:infall}

There is very little known about the existence, nature, and
composition of infall and/or outflow from our Galaxy.  Evidence for
infall is perhaps given by the observation of the High and Very High
Velocity clouds.  However, there is certainly a need for infalling
matter at early times, if one views infall to disk as just the outflow
from the halo as it collapsed.  In this case, one imagines a
metal-poor (i.e.~BBN composition) infall on a short ($\sim 2$ Gyr)
timescale.  Another justification for infall is more pragmatic: it
provides a possible solution to the G-dwarf problem (discussed in \S
\ref{sec:constraints}).

We will adopt models with and without infall, which we parameterize
according to the widely used form $f(t) = f_0 \exp(-t/\tau_{\rm
inf})$.  We will vary the infall strength $f_0$, and the timescale
$\tau_{\rm inf}$.  Namely, we will try $\tau_{\rm inf} = 2$, 4 Gyr,
and we will use different $f_0$ such that infall contributes 50\% and
99.9\% of the total disk mass today.  We will also consider ``closed
box'' models with no infall, i.e. for which $f_0 = 0$.

\subsubsection{Supernova Rates}

We will assume that all supernovae arise either from the core collapse
of massive stars (type II) or from accretion onto a CO white dwarf in
a binary system (type Ia).  All stars above $\sim 12 \msol$ are
thought to become type II supernovae, whereas the CO progenitors of
type Ia supernovae are arise from binary systems of intermediate
($3-8$ \msol) mass.  The rates for these events are given by
(Matteucci \& Greggio \cite{mg}), and are of the form
\beq
R_{\rm Ia}(t) = \lambda \, {\cal F}[\phi,\psi,f]
\eeq
where $\cal F$ is a functional of the IMF and SFR, as well as $f$, the
primary-to-secondary mass distribution in binaries (and perhaps also
the initial separation; e.g., Mathews et al.\ \cite{mbcs}).

The amplitude of the type Ia rate is controlled by the dimensionless
parameter $\lambda$ which is the probability for binary systems in the
appropriate mass range to undergo supernova events.  This number is
not well understood theoretically and in practice is adjusted to
reproduce the present ratio of type II to type Ia events $\sim 1$;
this ratio is proportional to $\lambda$ but is much larger due to the
larger number of low-mass, potential type Ia progenitors.  We will
examine the sensitivity of our results to this parameter, by choosing
two values for $\lambda$: a high value of $\lambda=0.05$ near the
value recommended by Matteucci \& Greggio (\cite{mg}), and a low value
$\lambda=0.007$ used by Timmes, Woosley, \& Weaver (\cite{tww}).

The type II nucleosynthesis yields are those discussed in \S
\ref{sec:mod-nuke}, and are directly included in eq.~(\ref{eq:dxidt}).
Type Ia yields of nuclide $i$ are added by putting
\beq
E_i^{\rm Ia} = M_i^{\rm ej} R_{\rm Ia}
\eeq
with $M_i^{\rm ej}$ the mass ejected in $i$ in a type Ia event.
We use the yields for the W7 model of Thielemann, Nomoto, \& Yokoi
(\cite{tny}).

\subsection{Constraints on the Global Models}
\label{sec:constraints}

We first constrain the global models to fit the solar neighborhood
properties within fairly generous tolerances, excluding for the moment
any consideration of the acceptability of the global models' light
element yields.  The available constraints on the gross features of
the models are $\sigma_{\rm g}^1$, $\mu_1$, and the CNOFe abundances.

Note that we want to be not too restrictive on our models, as the
data, even the solar system abundances, are likely to be subject to
systematic errors.  For example, Olive \& Schramm (\cite{os81})
suggest that the solar system may have been formed from a young OB
association; as a result its elemental and isotopic content would be a
biased indicator of the larger solar neighborhood.  Indeed, it seems
unwise to dwell on one constraint at the exclusion of the rest.
Consequently, we wish to be generous in choosing models with which to
check chemical evolution uncertainties in the light elements.

We will demand that all models fit the local gas mass fraction, which
is determined to be $\mu = 0.1 - 0.2$ (Rana \cite{rana}).
Specifically we require $\mu = 0.13-0.17$.  This fixes the overall
mass consumption.  Given that the a central feature explicitly built
into chemical evolution models is the mass consumption and the
conservation of overall mass, this is a most important constraint.  We
therefore have tuned all of the SFR normalizations to satisfy this
constraint.

As for the quantitative means of evaluating a model, the $\chi^2$
analysis of Tosi (\cite{tosi}) is interesting but perhaps premature
given that the data is likely to be fraught with systematic errors.
Instead, we will make the generous (and implementationally simple)
demand that the solar CNOFe abundances within a $\pm$0.4
dex\footnote{dex = base 10 logarithmic units; i.e.~a logarithmic
variation by $\pm$0.4 dex is a linear variation by a factor of
$10^{0.4}=2.5$} range.  This is to be compared with the factor of 2
that Timmes, Woosley, \& Weaver (\cite{tww}) allowed themselves in a
model tuned to give good results.

We will also constrain some of our models with observed G-dwarf
distribution.  G-dwarf stars are long lived, with ages comparable to
the Galactic age now.  Consequently, their provide a record of the
integrated star formation history of the galaxy; the metallicity
distribution of these stars is as seen in Figure \ref{fig:gdex}.  It
has long been known that the simplest models of galactic chemical
evolution are unable to reproduce the G-dwarf distribution.  In
particular, such models cannot avoid overproduction of low metallicity
stars, whereas the data shows a dearth of G-dwarfs below a metallicity
of [Fe/H].

One of the solutions to this classic ``G-dwarf problem'' is the
introduction of infall.  In models with infall, low metallicity stars
are still formed, but as material is added subsequently, the fraction
of low metallicity G-dwarfs becomes acceptably small.  We will thus
apply a G-dwarf constraint to our models, but only on models with
infall.  For these we will demand a solution to the most emphasized
part of the G-dwarf problem, namely the need for a small number of low
metallicity G-dwarfs.  We will ask that our models not exceed the
observed fraction of such stars at [Fe/H] = -0.8 by more than a factor
of 0.4 dex $\simeq 2.5$; this will prove to be a restrictive
condition.  We will not make the stronger demand that the models fit
the full G-dwarf distribution to a high accuracy, as this is not done
even in models which allow themselves tuning which we lack.

We will not impose the G-dwarf constraint on models without infall, as
it has long been thought (e.g.~Tinsley \cite{tins}; Pagel
\cite{pagelgd}) that such closed box models are unable to meet this
constraint without further modification.  On the other hand, Mathews
\& Bazan (\cite{matbaz}) show that the G-dwarf problem is somewhat
mollified by the inclusion of metallicity dependent stellar ages, and
Sommer-Larsen (\cite{s-l}) has noted a selection bias in favor of high
metallicity dwarfs.  At any rate, we keep the closed box models since
their ubiquitous presence in the literature demands that one examine
their results, and again, we want to err on the side of inclusiveness
in our selection of viable models.

\subsection{Light Element Model Features}

Having assembled a set of possible models and determined constraints
on their gross characteristics, we now turn to the model features we
will use to encompass the possibilities for light element evolution.
For the most part, these introduce small enough changes in the CNOFe
abundances that each model's CNOFe prediction will remain valid
despite the modification of light element evolutionary features.  An
exception to this rule comes from mass loss.

The evolution of all light elements is affected by their initial BBN
abundance as set by the choice of $\eta$.  We use three different
values of $\eta$, guided by the analysis of Copi, Schramm, \& Turner
\cite{cst} (cf.~eq.~\ref{eq:etalim}).  We choose central, high, and
low values $\eta = (2, 4, 6)\times 10^{-10}$; these values bracket the
Copi, Schramm, \& Turner's ``sensible'' limit.  The low limit is
intentionally chosen to be slightly below their recommended value (but
above their ``extreme'' lower limit), in order to allow comment on
recent speculation of the possibility of a low $\eta$.

In addition to systematically varying $\eta$, we will include the
following features for each element:

\subsubsection{D}

As we have discussed, D evolution is uniquely simple because there are
no stellar sources of D; the range in D results thus stems only from
the choice of $\eta$ and from global model features.

\subsubsection{\he3}

For high mass \he3 yields we used the results of Weaver \& Woosley
(\cite{wwprep}), which span a grid of masses from $\sim 11-40$ \msol\
and very complete range of metallicity from $Z=0$ to $Z=Z_\odot$.  The
metallicity dependence of these models is strong and thus a crucial
feature to include; these results update the work of Dearborn,
Schramm, \& Steigman (\cite{dss}).

As noted in \S \ref{sec:he3}, \he3 evolution hinges on the effect of
low mass stars on \he3.  To investigate this effect we have chosen an
approach similar to that of Vangioni-Flam, Olive, \& Prantzos
(\cite{vop}), as well as Olive et al.~(\cite{orstv}).  Namely, we
specify the \he3 survival fraction $g_3$ at $m = 1$, 2, and 3 \msol.
We choose to allow for: (1) modest \he3 destruction (Vangioni-Flam,
Olive, \& Prantzos found they must); (2) ``break even,'' with all \he3
surviving but no net production; and (3) modest \he3 production.
Specifically, the three ranges we choose are
\beqar
g_3\left(1\msol,2\msol,3\msol\right) & =  &\left\{
\begin{array}{l}
(1,0.7,0.7) \\
(1,1,1) \\
(1.33,1.33,1.33)\\
\end{array}
\right.
\eeqar
We chose the production to be small (i.e.~somewhat less that the Iben
\& Truran (\cite{it}) yields, corrected to include initial D, would
suggest) in order to be conservative.  It is clear, as pointed out in
Olive et al.\ (\cite{orstv}) that the yields at face value do not
work, assuming the conventional \he3 high mass yields.  In an effort
to allow for uncertainties in the Iben \& Truran (\cite{it})
calculation, we wish to see if even a modest production can work.

\subsubsection{\he4}

As discussed in \S \ref{sec:he4}, calculations of \he4 production must
allow for its production by stars having a wide range of masses;
because of this broad range, \he4 is particularly sensitive to
systematic problems in nucleosynthesis yields over the whole mass
range.  One such problems is that tabulations of stellar
nucleosynthesis yields usually do not include results for stars in the
range of roughly $8-12$ \msol.  There is not a firm understanding of
the basic nature of these stars' final fate; whether they become white
dwarfs or explode as supernovae is a delicate and unresolved question.
The nucleosynthesis results for these stars are therefore quite
uncertain, and indeed possibly quite different from those of stars
either more or less massive.  This is a regrettable state of affairs,
as a typical IMF places a good deal of weight on the yields from stars
in this range.  Thus whatever assumptions one adopts about these
yields can prove important for chemical evolution, and so we will
examine two possible scenarios.

Traditionally, one obtains yields in this mass range by interpolating
the results from the closest tabulated masses.  Such an interpolation
can be viewed as hedging one's bets about the accuracy of estimates
for the onset of core collapse, and we will include this case;
however, it has also been suggested that stars in this range produce
\he4 almost exclusively (Woosley \& Weaver \cite{wwrev}).  We
therefore mock up this behavior by assuming that {\it all} of the
ejecta from these stars is in this form.

Other important model uncertainties regarding \he4 production arise
due to the omission of mass loss in high mass stellar models
Consequently, we use the stated Woosley \& Weaver (\cite{ww93}) high
mass yields, which do not include the effects of mass loss, but we
also characterize the effects of mass loss by adding Maeder's
(\cite{mae92}) calculation of these.  Note that while Maeder also
calculated supernova yields which included mass loss, the models were
far less detailed and thorough than those of Woosley \& Weaver; thus
we choose to import only the mass loss addition.

\section{Results}
\label{sec:result}

\subsection{Effect of the Chemical Evolution Model Features}

The effect of our global model features are for the most part well
documented in the literature; we will discuss only the highlights.
For a more complete discussion we refer the reader to, e.g., Talbot \&
Arnett (\cite{ta}), Tinsley (\cite{tins}), and Tosi (\cite{tosi}).
Note that unless otherwise indicated, all of the figures in this
section are for models which vary only the parameters discussed while
keeping the others constant.

A sample of SFR behaviors appears in Figure \ref{fig:sfr}; Figure
\ref{fig:sfr} (a) displays the case of a closed box model.  We see
that for SFRs of the form $\psi = \nu \sigma_{\rm tot}\mu^n \propto
\sg^n$, there are progressively higher initial bursts and more rapid
declines for progressively higher $n$.  The presence of infall changes
the behavior of the SFRs whose form depends upon the total and/or gas
masses.  As seen in Figure \ref{fig:sfr} (b), for $\psi = \nu
\sigma_{\rm tot}$ the SFR is monotonically increasing with time as the
disk mass grows through infall, in contrast to the decaying SFR forms
in a closed box.  For $\psi = \nu \sg$ (Figure \ref{fig:sfr} (c)), in
the case of infall the SFR initially rises to a peak, but then falls
off as the gas is converted to stars and remnants.

The total and gas mass evolution is illustrated in Figure
\ref{fig:massev}. Our SFR boundary conditions and infall prescription
fix the $\sigma_{\rm tot}$ plot as well as the endpoints of the \sg\
plot.  Note that the spread in \sg\ at $t_1 = 15$ Gyr is an indication
of this spread in $\mu_1$, which arises due to slight differences in
the SFR tuning for different models.  The effect of infall on the gas
and total mass is clear from Figure \ref{fig:massev}.  Without infall,
the total mass is of course constant, and the gas mass monotonically
decreases.  With infall, the total mass rises rapidly, with the gas
mass initially following, then reaching a peak and turning over.

The relation between iron abundance and time (the ``age-metallicity
relation'') is an important one: iron is readily observable whereas
$t$ is not, and the maximal binding energy of Fe ensures that its
abundance will increase monotonically with time.  Thus the iron
abundance is traditionally used as a chronometer (although its origin
in both type Ia and type II supernovae makes it a somewhat problematic
one, with O perhaps being a more suitable choice.  See Wheeler,
Sneden, \& Truran \cite{wst}).

In Figure \ref{fig:fets} we plot [Fe/H] versus $t$ for all of our SFR
options, with data from Edvardsson et al.\ (\cite{edetal}).  In
general, we see that the Fe rises very quickly, with most of the
production happening at very early times and very little new Fe being
added today.  In comparing the curves in Figure \ref{fig:fets} to the
data it is immediately clear that all of the models are viable.  There
is also a great deal of scatter in the data, indeed, as much as the
overall trend.  Edvarsson et al.\ (\cite{edetal}) emphasize that this
scatter is real, and is not well explained in most chemical evolution
models which invoke, as we have, the assumption of instantaneous
mixing of stellar ejecta at a given epoch.  We can only expect to
reproduce the average trends in the observations, and one should
recall that individual stars can show considerable excursions from the
average trend.

The evolution of deuterium provides a good example of the effect of
different IMFs; the D history for three different IMFs is displayed in
Figure \ref{fig:imfs} (a).  The three choices have different return
fractions, with $R = 0.32, 0.42$, and 0.57.  As expected
(cf.~eq.~\ref{eq:dev}), the D evolution is very sensitive to this
parameter, with the overall depletion at the solar system (i.e.~at
$t_\odot = 10.4$ Gyr) varying from a factor of about 1.4 to a factor
of 2.3.  As we will see, this spread accounts for a large portion of
the model uncertainty in the D evolution (for a fixed $\eta$).  Note
that the largest D depletion comes from using the Scalo (\cite{scalo})
IMF, which is derived from the present-day mass function.  This IMF is
bimodal and so has a lot of power in the heaviest stars which eject
most of their mass; consequently, this yields a large return fraction
and considerable D depletion.

In Figure \ref{fig:imfs} (b) we show \he3 and D+\he3 for the same
models.  The tradeoff between D and \he3 is clearly seen, with \he3
being progressively higher in models having progressively lower D.
Indeed, despite the large variation in the individual D and \he3 cases
for the three IMF choices, we see that all show a very similar---and
nearly constant---evolution of D+\he3.  In Figure \ref{fig:imfs} (c)
we show the \he4 evolution for the same models.  These models show a
correlation between the D and \he4 evolution, as the models with
larger D depletion also have larger \he4 production.  We will
investigate this relationship in detail in \S \ref{sec:corr}.

Deuterium together with \he3 provides a good barometer of the effects
of infall on the evolution of an isotope.  The infall of primordial
material enriches the depleted Galactic D, while it dilutes the
increasing Galactic \he3.  This behavior is evident in Figure
\ref{fig:dheinf} which shows the slower decline of D in the infall
models, mirrored by the inhibited growth of \he3.  Note, however, that
in the case of D the spread (at the time of the formation of the solar
system) due to different treatments of infall is only about 15\%, much
less than that caused by different IMF choices.

The different behaviors of nitrogen in our models are illustrated in
Figure \ref{fig:yno} (a).  We focus on N behavior at the same low
metallicities which are observed extragalactic \hii\ regions, and we
include data from Pagel et al.\ (\cite{pageletal}) and Skillman et
al.\ (\cite{skiletal}).  While there is a large scatter in the data,
the trend seems better fit by the models with mixed and the purely
primary N{}. Note in particular that the mixed model does a
surprisingly good job at the low metallicities.  While the fit is to
somewhat an accident of our particular choices for the
primary/secondary N mix, it is amusing that the lowest metallicity
region---which dominates the \yp fit---comes out as well as it does.

The effect of secondary N on the $Y-$N relation is shown in Figure
\ref{fig:yno} (b), with the $Y-$O relation in Figure \ref{fig:yno}
(c).  For these plots we have chosen models with elevated \he4
production, as this provides a better fit to the data, (\S
\ref{sec:speclite}).  The models with pure secondary N do indeed have
a sharp increase at low metallicities; this arises from 
helium production in concert with suppressed N production due to the
dearth of its seed nucleus C.  Note, however, that the low metallicity
$Y-$N relation is much more gentle for the case of partial primary and
partial secondary N, the case that best fits the observations.  The
small departure from linearity in this case suggests that linear $Y-$N
fits might not be in much error (\S \ref{sec:speclite}).

\subsection{Global Constraints and Model Uncertainties}

Having sketched some of the effects of the global model features, we
now explore the range of abundances calculated for the initial suite
of candidate global models and for the subsequent set of light element
models.  In finding the range of light element predictions we obtain a
rough quantitative estimate for the model uncertainties inherent in
the chemical evolution arguments used to deduce the primordial
abundances of these elements.

We created the suite of candidate global models by making every
allowed combination of features shown in table \ref{tab:gmodf}; this
produces 1184 models.  We ran all of these models, using one
particular choice of light element model features.  Results of this
run are summarized in Figure \ref{fig:run1} (a), in which we plot the
abundances for each element, calculated at the birth of the solar
system ($t_\odot$, 4.6 Gyr ago).  Abundances are expressed in terms of
the logarithm with respect to the observed solar abundance ${\rm
A/H}_\odot^{\rm obs}$:
\beq
\label{eq:overab}
{\rm [A/H]} = \log \frac{{\rm A/H}(t_\odot)}{{\rm A/H}_\odot^{\rm obs}}
\eeq
(solar data is from Geiss (\cite{geiss}) for D and \he3, and from
Anders \& Grevesse (\cite{ag}) for all other elements).  Note the wide
range of all calculated abundances, including those for the light
elements.  Indeed, the only element without large excursions is \he4,
since it is the only element whose solar system abundance is mostly
primordial and so on this scale does not show large effects of
uncertainty due to chemical evolution.

We now impose the constraint that our models reproduce the solar
abundances of CNOFe within $\pm 0.4$ dex, i.e.~within a factor $\sim
2.5$.  While this may seem generous, recall that we have not tuned our
models to fit any elements in particular, yet even models that do so
(e.g.~Timmes, Woosley, \& Weaver \cite{tww}) allow themselves a factor
of 2.  Also, the scatter in the age--metallicity relation suggests
that there is a large spread of metallicities at a given time.  We
must be aware of the possibility that the solar data will not
necessarily reflect the average at the solar neighborhood, so we do
not want to exclude potentially good models on the basis of a single
element.

When we impose this constraint on the candidate global models, we
reduce the field to 267 models.  A plot of the solar abundances for
these models appears in Figure \ref{fig:run1} (b).  The light element
uncertainties are seen to be greatly reduced, while still being
considerable.  The spread in D and \he3 is of particular interest, as
we have not added any model variance in the D or \he3 evolution
(e.g.~different $\eta$ or different low mass \he3 stellar processing).
Thus the range abundances of D and \he3 reflect only the variance due
to the global models.  Quantitatively, we have a variance in D of a
factor of 2.0.  As we have noted, this is very close to the range due
solely to the IMF, which we see to be the largest source of D
variation.  We also find \he3 to vary by a factor of 1.9, and $\Delta
Y$ to vary from 0.018 to 0.046.

As discussed in \S \ref{sec:infall}, we also constrain the models with
infall to fit the G-dwarf distribution, again with a loose tolerance.
We focus on the ability the models to avoid a large number of low
metallicity G-dwarfs, comparing our results to the data of Rana
(\cite{rana}).  A plot of the distributions for one model that passes
and one that fails can be found in Figure \ref{fig:gdex}.  As
discussed in \S \ref{sec:infall}, we do not apply this constraint to
models without infall.  With the constraint, we reduce the number of
models to 114, with 28 of these being infall models.\footnote{ The
only single features clearly important in determining whether a given
model would pass or fail were the IMF slope and mass limits.  We found
no models with $x = 2$ survived; for $x = 2.35$, no models with the
widest mass limits passed; for $x = 2.7$, mostly the models with the
smallest mass limits passed.  Aside from the IMF, no other model
feature was prominent in either its presence or absence from the
passing models.  }

We now have a set of 114 chemical models which fit the solar system
behavior within the tolerances we have set.  To investigate the range
of light element evolution allowed by these models, we run each of
them for 18 different varieties of light element evolution models.
Note that this is far less than the possible number of allowed
combinations of options for light element evolution.  However, to run
such a number of models would not only be tremendously time consuming,
but it would also be redundant.  We do not expect, for example, the
various LiBeB yields in the cosmic ray models to depend on $\eta$ in
an important way compared to the other large uncertainties.  Thus we
have reduced the number of light element variations to only those 18
most likely to give interesting results.  In total, therefore, we have
18 light element models $\times$ 114 global models = 2052 overall
models to run.

Results for the 2052 models appear in Figure \ref{fig:run2-nocut}.
Notice that a problem has developed in the CNO abundances---some light
element evolution features have affected them.  The problem arises due
to the extra CNO contribution from massive stars with mass loss.  For
consistency with our protocol of constraints, we again constrain the
models to give CNOFe within $\pm 0.4$ dex of solar.

This reconstraint reduces the number of models by about 25\% to 1460.
Results appear in Figure \ref{fig:run2-cut} (a).  Note that even for
this smaller number of models, the range in the light element
abundances remains very similar, although with fewer models the
extreme variations are less populated.  The models depicted in Figure
\ref{fig:run2-cut} (a) will be the basis for the analysis in the
sections below, except where otherwise noted; we will refer to these
as the ``final set'' of models.

The source of the gap in the D abundances in Figure \ref{fig:run2-cut}
(a) becomes clear when we plot the results for a particular $\eta$.
As Figure \ref{fig:run2-cut} (b) shows, the D gap arises due to the
very high initial D at $\eta = 2 \times 10^{-10}$.  It is also clear
that it is hard for our models to reduce such a high initial D to its
presolar abundance, a point we will explore in more detail in the next
section.  Note that \he3 also shows sensitivity to its initial
abundance.

The spread in the light element abundances, particularly the spread
beyond that in the ``control'' group CNOFe, gives a sense of the
spread due to model dependence.  Note, however, that one must be
careful in interpreting these limits.  While we have tried a broad
sample of models, the lack of a firm theory of star formation leaves
open the possibility that one might construct other ones.  Moreover,
recall that we have not used chemical evolution to get limits on the
primordial abundances.  Instead, we have assumed BBN yields and seen
how well these fit the solar system.  What one really would like to
know is not this but rather the opposite case.  Namely, one would
construct models with variable primordial light element abundances
(i.e.~not necessarily given by BBN values at a given $\eta$), and
constrain these models to reproduce the solar system values for the
light elements and everything else.  The resulting spread in the
primordial abundances would be an indication of the model
uncertainties incurred when setting limits on $\eta$.  This procedure
suffers from an exclusive reliance on the veracity of chemical
evolution models.  Furthermore, we may estimate the model ranges such
a calculation would give, as we now discuss.

Table \ref{tab:evar} summarizes the model variations in the light
element production for each $\eta$.  We see that, for all $\eta$, D
varies almost exactly by a factor of 2 and $\Delta Y$ varies by
$0.063$ units.  The variation in D+\he3 depends somewhat on $\eta$,
but is small, varying by $< 20\%$ from $\sim 2.8$.  The
$\eta$-independence of these quantities are evidence that the range
model variations does not depend on the light element initial values.
This gives us confidence that, were we to do a light element evolution
study by allowing the primordial values to float (as described in the
preceding paragraph), we would find similar ranges.  That is, we
expect the reduced quantities in table \ref{tab:evar}, namely D$_{\rm
max}$/D$_{\rm min}$, (D+\he3)$_{\rm max}$/(D+\he3)$_{\rm min}$, and
$Y_{\rm max}-Y_{\rm min}$, to be generic estimates for the chemical
evolution model ranges of these elements.  Of course other selections
of model features will give different numbers, but these results at
least give a quantitative estimate for the model dependence within a
large scale of models, and serve as a benchmark for others.

We emphasize that the D variation is due solely to variation of global
model features, in particular the IMF and also infall.  Thus to reduce
the variation in D requires better chemical evolution modeling as a
whole.  By contrast, the span in the \he3 and \he4 ranges are due in
part to the effects of different stellar processing features for these
elements.  Improvements in these features could reduce the ranges
given, independently of improvements in the global chemical evolution
framework.

It is encouraging to note that the variations in D, \he3, and \he4,
while large, are significantly smaller than the 0.8 dex (i.e.~factor
of 6.3) overall variations allowed for CNOFe in our global model
constraints.  This implies that the light elements through \he4 have
less model uncertainty than do the heavy elements.  However, the
relationship between the heavy and light element uncertainties is not
necessarily a simple one.

To get a feel for the connection between the heavy and light element
model variations, we examine the effect of tightening the fiducial
tolerances in the heavy elements.  Specifically, we now require that
our models reproduce solar CNOFe to within $\pm 0.3$ dex, i.e.~to
within a factor of 2.  Results appear in Figure \ref{fig:run2-bigcut}.
The reduction in the number of models is drastic, going from 1460
models at $\pm 0.4$ dex tolerance to 479 models at $\pm 0.3$ dex
tolerance, a 66\% reduction in models from only a 20\% tightening of
the constraint.  One can understand this large reduction by noting
that our heavy element yields generically are scattered around the
solar levels.  For example \c12\ is produced on ``average'' above the
solar level, while \o16 is produced in a large range centered below
solar; in other words, the O/C ratio is typically subsolar.  Thus our
models have some range of error in reproducing O/C$_\odot$, and will
never have all abundances any more accurate than this range.  Indeed,
demanding simultaneous fitting of both abundances for progressively
smaller tolerances would eventually eliminate all models when the
constraints become too stringent given the heavy element yields we
have used.

Despite the large reduction in number of models which survive the
stronger heavy element constraint, the spreads in the light element
abundances remain roughly the same.  This reinforces the notion that
the CNOFe spreads are to some degree decoupled with the light element
spreads.  More quantitatively, we find for these models that D varies
by about a factor of 1.76 and \he3 by a factor of 3.1, values closer
to the range for models with the $\pm$0.4 dex CNOFe tolerance than the
20\% tightening would naively allow.  In other words, by making the
variation in CNOFe smaller by 20\%, we do not reduce the variation in
D and \he3 by as large a factor.  Thus the model variation in the
heavy and the light elements are not related by a simple linear
scaling, but instead the relationship between the two seems to be more
subtile.

\subsection{Correlations in Light Element Evolution}
\label{sec:corr}

We now study in detail the results from the set of 1460 models
comprising the ``final set'' having all allowed light element
variation while also fitting CNOFe.  In particular, we will be
interested looking at D, \he3, and \he4 (pre)solar and ISM predictions
simultaneously to see better the correlations in their evolution and
to show the interplay of the observational constraints.

In Figure \ref{fig:dshe3s} we plot the presolar D abundance versus the
presolar \he3 abundance for each model.  One notices immediately that
the choice of $\eta$ (different shaped points) is the largest effect
controlling the presolar abundance of both elements.  Furthermore, the
correlation between D and \he3 is evident: the negative slope in the
lines traced out in the Figure points out the well-known tradeoff that
D destruction leads to \he3 production.  The steepness of these lines
(which radiate out from the primordial values at each $\eta$), depends
on the degree of low mass destruction or production of \he3.  The
highest of the three ``prongs'' for each $\eta$ comes from the case of
\he3 production in low mass stars.  The middle prong is the ``break
even'' case, and the lowest one from net \he3 destruction.  We clearly
see that the \he3 presolar yields are very sensitive to the low mass
stellar behavior, which we have only varied by 30\% either way from
break even.  Our models thus confirm the results of Vangioni-Flam,
Olive, \& Prantzos (\cite{vop}) and Olive et al.\ (\cite{orstv}), and
extend them to include a very wide range of models.

Figure \ref{fig:dshe3s} also includes lines marking the $2-\sigma$
variation in the observed presolar D and \he3 abundances, as
calculated by Geiss (\cite{geiss}).  The Figure provocatively shows
that none of the low-$\eta$, i.e.~high initial D and \he3, models are
able to fit the presolar data to within this tolerance.  Indeed, the
majority of the moderate-$\eta$ models also seem to fail.  At face
value, this might suggest that low values of $\eta$ are ruled out and
that higher ones are favored.  The question is whether this strong
conclusion holds up under scrutiny.

As we have discussed in the previous section, it is a subtle matter to
use our models to constrain the possible BBN abundances.  We have not
included all possible chemical evolution models (e.g.~IMF and SFR
schemes) and one cannot conceivably do this, given the freedom one has
to adjust each.  Also, recall the observed scatter in Fe abundances
for stars which are presumably coeval.  If the presolar D is a low
variation to the mean we have calculated, we would be incorrectly
constraining our results.  Finally, as we have noted, significant
uncertainties and assumptions underlie the entire chemical evolution
framework we have adopted (and variants thereof).  Even if one could
show that if no such models work to explain the low-$\eta$ D and \he3,
such a conclusion would not be founded upon a first principles
calculation, since none exists.

Nevertheless, it is not terribly surprising that $\eta = 2 \times
10^{-10}$ fails for our study.  Indeed, as we have noted, this value
is outside of the ``sensible'' range in $\eta$ given by Copi, Schramm,
\& Turner (\cite{cst}), although it is within the ``extreme'' range.
Also, while chemical evolution models are uncertain, as we have
argued, the D evolution is perhaps the most certain of any nuclide
considered.  Thus there is no obvious reason to disbelieve the D
results, particularly when other abundances can be fit reasonably
well.

Furthermore, while chemical evolution is known to be uncertain, our
results as presented help one to quantify the size of the model
dependence.  At $\eta = 2 \times 10^{-10}$, the D abundance varies by
a factor of $\sim 2$; but the lowest D abundance for this $\eta$ is
still a factor of $\sim 1.5$ away from upper limit to the solar system
observation (and with an undesirable \he3 abundance at that).  It is
incorrect to think of the spread in our results as some sort of
distribution around a mean model, as we have not just performed smooth
parameter variations, but we have also changed whole prescriptions
for, e.g., the SFR.  Thus it is hard to say more than that the
presolar value of D for BBN with $\eta = 2 \times 10^{-10}$ is not
reachable in our models and seems hard to reach by models similar to
ours.

While presolar D is a strong constraint on models with high $\eta$, we
see that presolar \he3 is a very strong constraint for all $\eta$.
Because \he3 survives processing of both D and \he3, its abundances
grows with time.\footnote{This is true for sufficiently early epochs
and for sufficiently high $g_3$, both of which are the case in our
study.}  However, the presolar \he3 is not much larger than its
primordial level, and so demands that models produce very little \he3.
We see that models with high $\eta$, i.e.~with the lowest initial
\he3, are the only ones that can fit the presolar constraint.  Within
these, the different \he3 processing becomes important, with \he3
production in low mass stars often overproducing \he3.  Since this low
mass behavior nevertheless seems to be demanded by planetary nebula
data, the data seems to favor models which have very little processing
in general.

We may also examine correlations between the calculated (pre)solar
abundances of other pairs of light elements.  Figure \ref{fig:dshe4s}
shows D and \he4 (pre)solar abundances.  Here again, the largest
segregation of points is due to the different initial D abundances for
different $\eta$.  The shapes of the point distributions are similar
for the different $\eta$ values, though the higher $\eta$ regions are
more compressed.  Note that the regions are shifted upwards for
increasing $\eta$, reflecting the higher \yp\ for these models.
However, it is clear that the shift due to different \yp\ values is a
much smaller effect than the spread due to chemical evolution effects,
which give $\Delta Y$ anywhere from 0.04 to 0.08.

For a given $\eta$, the D--\he4 correlations are not as immediately
clear as for the D--\he3 plot.  One can understand the trends by
separating the points according to the \he4 model features to be
varied, namely the $8-12$ \msol\ yields and the presence of mass loss
in massive stars.  For models with no extra \he4 yields (i.e.~using
just the adopted yields without mass loss and interpolating between
$8-10$ \msol), the predictions appear as the lowest band of points;
data for these models alone appears in Figure \ref{fig:dshe4s} (a).
These models do indeed show an appreciable correlation, as we have
argued they should: D depletion is the measure par excellence of the
amount of material processed through stars, and \he4 production is one
consequence of this processing.  Thus we expect and find a negative
correlation for the ``standard'' case.

Figure \ref{fig:dshe4s} (b) shows the models with one or both of the
helium enhancing features.  We see here, like the D-\he3 plot, there
are three ``prongs'' for each $\eta$.  These correspond (for the most
part), to the three different cases of ``nonstandard'' \he4 yields.
The lowest trend is for mass loss only, the intermediate (and some of
the highest) prong is due to high $8-10$ \msol\ yields, and the
highest points come from a combination of the two.  Here again, we see
the negative correlation we expect.  We also note that many models
appear to fit the solar \he4.  Indeed, while some are high and some
low, a good number of both standard and nonstandard \he4 yields give a
successful fit.

We should emphasize that the physics that underlies the D--\he4
correlation is different from that behind the D--\he3 correlation.  In
the latter case, we have demanded that D be made into \he3, and so we
have explicitly built in the correlation between the two.  The
relation between D and \he4 is somewhat less direct, as Galactic \he4
does not come from primordial D, but rather the destruction of D
happens in processes which also produce \he4 independently of the
level of D.  Finally, having found correlations between D and \he3 as
well as D and \he4, it one also expects correlations between \he3 and
\he4 production.  These exist but are more complicated at there is an
interplay of both the \he3 and \he4 evolutionary features.  The basic
trend, however, is unsurprising: a correlation between the two.

Having examined the correlations between different model calculations
at the epoch of solar birth, we now turn to correlations at the
present epoch.  Here the underlying physics is the same and so we of
course expect similar relationships between the model yields, but the
observational constraints are different, and in the case of D, more
severe.  In Figure \ref{fig:dihe3i} we plot D versus \he3 for the
present epoch, i.e.~we plot the abundances calculated for the ISM now.
One notices the same features as the plot for the solar system, now
smeared out with the intervening evolution.  For \he3 the
observational situation is uncertain as the data shows a large
dispersion; if one included the range of abundances spanned by the
observations it would not constrain any of these points.  This
contrasts with the power of the presolar \he3, which offers a tighter
constraint than the presolar D.

The D observations in Figure \ref{fig:dihe3i} are from the {\it HST}
observation of Linsky et al.\ (\cite{linsky}), with the $2-\sigma$
variation shown.  (Unfortunately, we are unaware of accurate \he4
abundances in the ISM, and so we are unable to constrain our models
with this isotope.)  If this number is to be taken seriously, it is a
strong constraint on chemical evolution models.  Here again, we find
that the low-$\eta$ points are very disfavored by the D observations,
with the lowest points in this regime needing to move at least a
factor of 2 to meet the data.  On the other hand, the spread in these
points is larger than the presolar data, being almost a factor of 3.
Recalling the caveats above, we note that our models have trouble
fitting the ISM D abundance given an initial D from a low-$\eta$ BBN
model.

The stringency in the observed presolar and ISM abundances for D,
illustrated in figures \ref{fig:dshe3s} and \ref{fig:dihe3i}, bears
further investigation.  As we have noted, D is the nuclide whose
chemical evolution is the simplest, and thus we can expect a given
model to calculate D evolution the most accurately of all the
nuclides.  It is thus interesting to ask what models survive the
constraint that both the presolar and ISM D observations are fit to
within $2-\sigma$.  As is clear from figures \ref{fig:dshe3s} and
\ref{fig:dihe3i}, the more stringent of the two cuts is comes from
fitting the ISM observation.

Demanding that models fit presolar and ISM D reduces the number of
models from 1460 to 231.  In Figure \ref{fig:dshe3sdcut} we plot the
calculated presolar D versus \he3 for these models.  Comparing these
results to those of Figure \ref{fig:dshe3s}, we see that some of the
models which could otherwise fit the (also stringent) presolar \he3
are removed.  However, many of these models still remain.
Encouragingly, most of the models which successfully fit the presolar
and ISM D also fit the (pre)solar \he3 and \he4.

While it is not our purpose to find the ``best models'' for light
element chemical evolution, it is useful to consider the reason for
the success of the models which fit the observed (pre)solar D, \he3,
and \he4 abundances, as well as the ISM D.  Most importantly, these
models all have a high $\eta$ and so begin with a very small amount of
D and \he3---indeed, at D/H$_{\rm p} = 2.3 \times 10^{-5}$, they are
already smaller than the $2-\sigma$ upper limit to the presolar
abundance.  Thus these demand very little stellar processing of D to
reduce it to the observed abundance, and therefore need to avoid
producing much \he3.

Interestingly, these models have very few other strong identifying
characteristics.  All low mass \he3 prescription are represented in an
even way, as are all combinations of \he4 prescriptions.  The global
characteristics include all SFR prescriptions and infall as well as
closed box models.  The two steepest IMF slopes are both represented
as well, although the only mass limit range is the most restricted one
(type 2 in table \ref{tab:gmodf}).

We are left to conclude that D evolution is a strong constraint on
chemical evolution models, but also that models meeting this
constraint can still fit \he3 and \he4.  We have demonstrated that
comprehensive evolution of the light elements is possible within our
adopted framework of chemical evolution, although this framework
allows for substantial variation in its results according to the
adopted model prescriptions and parameters.

\subsection{Additional Considerations}
\label{sec:speclite}

\subsubsection{D}
The program of observation of D in quasar absorption line systems, now
just beginning, is likely to revolutionize how BBN is observationally
constrained.  The power and potential importance and of this method
impel a close scrutiny of its assumptions.  In particular, it is
implicitly assumed that the observed D abundance in high redshift
systems is a direct record of the primordial abundance.  However, by a
redshift of $z \sim 3$, the universe it at least 1 Gyr old, and as
much as 3 Gyr old (the range comes about since the $t(z)$ relation
depends on $\Omega_0$).  Potentially, in this amount of time there
might be some depletion of D in the protogalaxies that comprise the
absorption line systems; the question is how large such depletion
could be.  Also, it is noted that typical absorption line systems are
not as metal poor as that studied by Songalia et al.\ (\cite{schr}),
Carswell et al.\ (\cite{crwcw}), and Tytler \& Fan (\cite{tf}); it is
thus desirable to understand the evolution of D as a function of
metallicity.

Regarding the depletion of D in the absorption line systems, we note
first that if these are protogalaxies, we would expect them to take
some time to form; however, we model the Galaxy once it has formed.
Thus an object at a universal age of 1 Gyr would correspond to a time
earlier than 1 Gyr on our models (indeed, Timmes, Lauroesch, \& Truran
(\cite{tlt}) argue for a significant delay before the onset of
galactic nucleosythesis).  And furthermore, a glance at Figure
\ref{fig:imfs} suggests that the D depletion at very early times is
minimal.

To address this question more systematically, we have considered the D
depletion in all of our models for universes at a redshift of $z=3$,
where we have made the very conservative assumption that the
protogalaxies form immediately and so time in our models corresponds
to universal time.  The correspondence between the observable $z$ and
the time $t$ in our models depends on the adopted cosmology, in
particular on the value of $\Omega$.  To be conservative we have
chosen $\Omega=0$ (i.e.~a curvature dominated universe, which has the
largest age at $z=3$), with a Hubble constant of $H_0 = 75$ km/s/Mpc.
For each of these cases we run our models to the time corresponding to
$z=3$ and find the D abundance.  To get a sense of how this correlates
with the metal production, we plot the D abundance for these models
against [Fe/H].  Results appear in Figure \ref{fig:dfez3o0}.  The
Figure confirms that to a high accuracy ($\ga 20$\% and usually better
than 5\%) D may be considered undepleted in these systems.
Furthermore, Figure \ref{fig:dfeimf} illustrates that for metal poor
systems with [Fe/H] $\la -1$, and not necessarily at $z=3$, the D
depletion is minimal regardless of the IMF and infall choices.

\subsubsection{\he4}

The results from the previous section make clear that our \he4 options
can produce very different \he4 abundances at the birth of the solar
system and in the ISM.  We now turn to the issue of how these features
affect \he4 at low metallicities, in particular how the model features
affect the $Y-$CNO relation studied in extragalactic \hii\ regions.
We show this relation for all four \he4 model options (all at
$\eta=4\times10^{-10}$) in Figure \ref{fig:ycnoop}.  We see that the
$8-10 \msol\ $ enhancement is effective at increasing the \he4 slope,
making an enhancement of about a factor of two over the standard case.
On the other hand, mass loss is unimportant in this regime.  This
result is not surprising when one considers that mass loss---at least
as we have modeled it following Maeder (\cite{mae92})---not only
increases \he4 but also CNO.  Thus while the \he4 production is
enhanced, the tracers are as well and so the net effect on $Y-$CNO is
small.  A note of caution, however, is that the mass loss calculations
are difficult and so the results we use are subject to large
uncertainty; also, they are only available for two metallicities.
Improved modeling is crucial to help address the issue of the $Y-$CNO
relation.

Figure \ref{fig:yno} (a) shows effect of secondary N evolution on the
low metallicity $Y-$N relation; this relation needs to be well
understood to properly extrapolate from the observations to derive the
primordial \he4 abundance.  Oftentimes a linear relation is assumed;
we now estimate the error in assuming a linear $Y-$N relation for
different cases of N evolution.  For each model, we will try to
reproduce the empirical fitting procedure by finding the slope of the
$Y-$N curve at a point and then extrapolating to get the estimated
primordial \he4 abundance $Y_{\rm p}^{\rm est}$.  We can then compare
this to the actual \yp\ value in the model and thus compute the error
$\delta Y_{\rm p}= Y_{\rm p}-Y_{\rm p}^{\rm est}$ in this
extrapolation procedure.

A potential complication to this procedure is that those models
lacking features which enhance \he4 (mass loss and enhanced $8-12
\msol\ $ production) will have smaller $Y-$N slopes and so will be
poor approximations to the extragalactic \hii\ region data.  Thus we
will also keep track of the average slope one would assume for these
regions if one did a linear fit; only the results with a sufficiently
high slope will be admissible for this comparison.

We perform the extrapolation at N/H $= 50 \times 10^{-7}$, i.e.~at a
region surrounded by data and not at low enough metallicity to itself
betray the low-N dropoff in \he4.  The average slope is not the one
used in the extrapolation but is taken at N/H $= 100 \times 10^{-7}$.
This quantity serves only as a diagnostic to indicate the goodness of
fit of the given model's $Y-$N relation to the observed slope.  As
such, a central point better quantifies the average behavior over the
whole region.

In Figure \ref{fig:yerror} we plot the error $\delta Y_{\rm p}$ in the
\he4 linear extrapolation versus the average slope $dY/d$N one would
assume for the region.  Note the distribution in slopes; we will only
consider models with slopes above $10\times 10^{4}$, which is a rough
lower limit to the slopes derived from the \hii\ region observations.
For the models with large slopes, one can immediately discern the
trends seen in Figure \ref{fig:yno}, here writ large.  Namely, the
models with no secondary N are well fit to a line, showing little
error in the extrapolation.  Indeed, the errors are fairly evenly
spread around zero, with a width of about 0.001 units in the mass
fraction.  The models with some secondary and some primary N give
$\delta Y \sim 0.002 - 0.004$, while models with purely secondary N
can have corrections as large as 0.015.

Thus the models with exclusively secondary N give $\delta Y_{\rm p}$
at the level claimed by Fuller, Boyd, \& Kalen (\cite{fbk}), Mathews,
Boyd, \& Fuller (\cite{mbf}), and Balbes, Boyd, \& Mathews
(\cite{bbm}). However, as pointed out by Olive, Steigman, \& Walker
(\cite{osw}), the N--O relation in extragalactic \hii\ regions demands
some degree of primary N; indeed, Figure \ref{fig:yno} fits the data
well with a mixture of primary and secondary N.  For this case, the
error in \yp\ extrapolation is small.  Indeed, differences at this
level are unimportant compared to the $\delta Y \simeq 0.008$
observational uncertainty.

It is amusing to note that linear extrapolations of \hii\ region data
often give slightly larger values when using N as a tracer than when
using O, with the difference being around 0.002 to 0.003 units, just
the level of the effect we see in the primary + secondary N models.
Clearly more work on this issue is needed, and the good agreement of
our mixed N model may be somewhat fortuitous, as we have not
explicitly built models of the irregular galaxies in which these
observations are made.  Nevertheless, we our models suggest that the
the effect of secondary N on \he4 extrapolation may not be as
pronounced as originally suggested.

\section{Conclusions}

Our results on the sources of our model sensitivity confirm those of
Tosi (\cite{tosi}), as we find that the elemental abundances of our
models are most sensitive to the shape and mass range of the IMF.
Indeed, the models we tried were incompatible with a very shallow IMF
slope for any mass range; this was the only model feature that was
completely excluded by our solar system constraints.  Furthermore, we
find that our ``best'' models, those fitting the solar CNOFe as well
as solar and ISM observations of D, select an intermediate IMF slope,
namely the Salpeter (\cite{salp}) value.  We also find significant
sensitivity to the presence and timescale of infall.  However, infall
itself is strongly constrained by the low metallicity G-dwarf
distribution.  Finally, we do not find our results to depend strongly
on the choice of the star formation rate.

We also follow Tosi (\cite{tosi}) in finding that our results are very
sensitive to the stellar nucleosynthesis yields employed.  We
emphasize in particular the effect of systematic effects such as mass
loss which are not included in many models.  In our models mass loss
can be an important source of \he4 as well as CNO; more detailed
calculations of this affect are needed to address the important
question of the \he4 contribution of high mass stars.  We also find
that our results are sensitive to the behavior of stars in the poorly
understood $8-12$ \msol range; this too provides an important
uncertainty for \he4 production.  Work on these objects would be as
useful as it will be difficult.  Finally, we stress the need for a
good understanding of the \he3 yields in stars of all masses.  While
it would be quite useful to update the Iben \& Truran (\cite{it})
calculation of \he3 production in low mass stars, it is as important
to understand \he3 behavior in high mass stars, with emphasis on the
effect of winds.

Regarding the viability of light element chemical evolution, it is an
important point simply that we do find models which can fit solar and
ISM constraints on D, \he3 and \he4.  This important ``existence
proof'' demonstrates that the framework we have adopted, despite its
roughness, is able to fit the data.  Thus a consistent picture of
light element evolution, from the early universe to the present day,
can be drawn.

{}From the perspective of BBN, it is reassuring that we find the effect
of different $\eta$ so important for D and \he3 chemical evolution.
We see that the global chemical evolutionary models we have examined,
while having a significant degree of model dependence, nevertheless
preserve information about the primordial abundances.  Thus we are
encouraged in our use of solar and ISM data as meaningful constraints
on BBN, with the caveat that there remain significant uncertainties
arising from chemical evolution.  Indeed, as we have shown, for D and
\he3 the model dependences are larger than the observational
uncertainties (for solar D and \he3 as well as ISM D).  Thus BBN
constraints coming from the solar and ISM D and \he3 are useful but
should be regarded with caution.

On the other hand, we do not find D to be subject to significant
depletion or chemical evolutionary uncertainty at early epochs probed
by measurements of quasar absorption line systems.  We find the D
abundances at this epoch to be a clean indication of the initial D
abundance (and thus of $\eta$); this result is independent of the
particular cosmology chosen.  Thus more observations of these
systems\footnote{Note however that deriving D abudances from the
observations may not be simple (Levshakov \& Takahara \cite{lt})}
should eventually be able to put strong constraints on $\eta$ and thus
on $\Omega_{\rm B}$.

In contrast to the case of D and \he3, we find \he4 evolution to be
dominantly sensitive not to its initial abundance but to the
uncertainties in its stellar production.  However, we do find that
within these uncertainties it is possible to find \he4 evolution at
low metallicity which can fit the $\Delta Y/\Delta$N and $\Delta
Y/\Delta$O slopes better than standard models would indicate.
Further, we find that the low metallicity $Y-$N trend is indeed
sensitive to the prescence of secondary N.  However, our models which
best fit \hii\ region N and O data require some degree of primary N
and consequently do not lead to large deviations from linear $Y-$N
relations.  This issue, however, merits further study.

Future work in light element chemical evolution in part would refine
the framework adopted here, most importantly by the adoption of
different models of the halo phase.  One might eventually hope to
create a self-consistent stellar and chemical evolution scheme for the
light elements, along the lines of the approach of Timmes, Woosley, \&
Weaver (\cite{tww}).  However, there is ultimately a need for a more
fundamental understanding of star formation and its relation to the
dynamics and chemistry of the Galaxy.  In particular, an understanding
of the physics behind the SFR and the IMF would be of tremendous
value.  Nevertheless, while uncertainties in chemical evolution
remain, it is encouraging that they still allow models for the light
elements to teach us about about BBN.

\acknowledgments{I thank Dave Schramm for his thoughtful advice on all aspects
of this paper.  I am grateful to acknowledge Frank Timmes for his
support and insight, and Jim Truran for helpful discussions.
Thanks to Keith Olive, Grant
Mathews, and Mike Turner for useful to discussions and to Jeff Harvey
and Simon Swordy for their time and energy.  Finally, it is a pleasure
to acknowledge the great technical advice of Craig Copi, and
the careful reading and pertinent comments of the referee,
Grant Bazan.
Presented
as a thesis to the Department of Physics, the University of Chicago,
in partial fulfillment of the requirements for the Ph.D. degree.  This
work was supported by NASA through a GSRP fellowship,
by NASA grants NGT-50939 and NAGW 1321,
by the National Science Foundation grant NSF AST 94-20759, and
by the Department of Energy grant DOE DE GF02 91ER 40606.}

\clearpage


\clearpage

\begin{table}[ht]
\begin{center}
\caption{Galactic Sources for the Light Elements}
\label{tab:source}
\begin{tabular}{clc}
\hline \hline
{\sc Nuclide} & {\sc Production site} & {\sc Contribution to} \\
  & & {\sc observed abundance} \\
\hline
D & none & not applicable \\
\he3 & low mass ($\la 3 \msol$) stars (?) & $\ga 50 \%$ (?)$^*$\\
\he4 & stars of all masses & $\sim 10-20 \%$ \\
\hline
\end{tabular}

$^*$ \he3 abundances in the ISM vary with mass of \hii\ region
\end{center}
\end{table}

\clearpage

\begin{table}[h]
\begin{center}
\small
\caption{ Global Model Features}
\label{tab:gmodf}
\begin{tabular}{ccl}
\hline \hline
Model Feature & \# & Descrption \\
\hline
SFR & 1 & $\nu \sigma_{\rm tot}$ \\
    & 2 & $\nu \sg$ \\
    & 3 & $\nu \sigma_{\rm tot}\mu^2$ \\
    & 4 & $a \exp(-t/\tau)$, $\tau = 7.5$ Gyr \\
    & 5 & \hphantom{$a \exp(-t/\tau)$, } $\tau = 15$ Gyr \\
IMF shape & 1 & $\phi \propto m^{-(1+x)}$ \\
          & 2 & from PDMF (Scalo 1986)$^*$ \\
IMF slope & 1 & $x = 1.0$ \\
          & 2 & $x = 1.35$ \\
          & 3 & $x = 1.7$ \\
IMF mass limits & 1 & $(m_l,m_u)$ = (0.2,100) \\
                & 2 & $(m_l,m_u)$ = (0.1,60) \\
                & 3 & $(m_l,m_u)$ = (0.4,30) \\
infall &   & $f = f_0 \exp(-t/\tau_{\rm inf})$ \\
       & 1 &  $f_0 = 0$ \\
       & 2 &  $\tau_{\rm inf} = 2$ Gyr; 99.9\% $\sigma_{tot}$ \\
       & 3 &  $\tau_{\rm inf} = 4$ Gyr; 50\% $\sigma_{tot}$ \\
       & 4 &  $\tau_{\rm inf} = 4$ Gyr; 99.9\% $\sigma_{tot}$ \\
N yields &   &
    $X^{\rm ej} = X_0^{\rm ej}(\alpha + \beta (X_C/X_C^\odot))$ \\
         & 1 & $\alpha = 1$, $\beta = 0$ \\
         & 2 & $\alpha = 0.5$, $\beta = 1$ \\
         & 3 & $\alpha = 0$, $\beta = 1$ \\
         & 4 & $\alpha = 0$, $\beta = 2$ \\
SN Ia normalization & 1 & $\lambda = 0.05$ \\
           & 2 & $\lambda = 0.007$ \\
\hline
\end{tabular}
\end{center}
\centerline{$^*$Only calculable for exponential SFR}
\end{table}

\clearpage

\begin{table}[htb]
\begin{center}
\caption{{Model Ranges for Solar D, $^3$He, and $^4$He Production}}
\label{tab:evar}
\begin{tabular*}{6.0in}{c@{\extracolsep{\fill}}c@{\extracolsep{\fill}}
     c@{\extracolsep{\fill}}c@{\extracolsep{\fill}}
     c@{\extracolsep{\fill}}c@{\extracolsep{\fill}}c}
\hline \hline
{$\eta$} & $10^5$ D$_{\rm min}$ & $10^5$ D$_{\rm max}$ &
$10^5$ \he3$_{\rm min}$ & $10^5$ \he3$_{\rm max}$
& $Y_{\rm min}$ &  $Y_{\rm max}$ \\
\hline
$2 \times 10^{-10}$ & 5.90 & 11.9 & 4.26 & 18.9 &
0.250 & 0.313  \\
$4 \times 10^{-10}$ & 1.82 & 3.66 & 2.00 & 7.25 &
0.258 & 0.322  \\
$6 \times 10^{-10}$ & 0.912 & 1.83 & 1.44 & 4.26 &
0.262 & 0.326 \\
\hline
\end{tabular*}
\begin{tabular*}{6.0in}{c@{\extracolsep{\fill}}c@{\extracolsep{\fill}}
     c@{\extracolsep{\fill}}c}
$\eta$ & D$_{\rm max}$/D$_{\rm min}$ &
     (D+\he3)$_{\rm max}$/(D+\he3)$_{\rm min}$
  & $Y_{\rm max}-Y_{\rm min}$ \\
\hline
\hline
$2 \times 10^{-10}$ & 2.02 & 3.03 & 0.063 \\
$4 \times 10^{-10}$ & 2.01 & 2.86 & 0.064 \\
$6 \times 10^{-10}$ & 2.01 & 2.58 & 0.064 \\
\hline
\end{tabular*}
\end{center}
\end{table}

\clearpage

\begin{center}
{\bf FIGURE CAPTIONS}

\begin{enumerate}

\item
\label{fig:schplt}
{Big bang nucleosynthesis yields as a function of the baryon-to-photon
ratio $\eta$.  The \he4 abudance is plotted as a mass fraction $Y_{\rm
p}$; all other elements given as number relative to hydrogen.  Open
points show abundance used for the three values of $\eta$ considered
in this study.  The dashed lines give the bounds for the Copi, Schramm
\& Turner (\protect\cite{st}) ``sensible'' bounds on $\eta$.}

\item
\label{fig:sfr}
{(a)The different adopted star formation rates, plotted as a function
of time for models without infall.  (b) The star formation rate $\psi
= \nu \sigma_{\rm tot}$ plotted for the different models of infall.
(c) As in (b), for $\psi = \nu \sigma_{\rm tot} \mu^2$.}

\item
\label{fig:massev}
{The (a) total and (b) gas mass surface densities as a function of
time.  Plotted for $\psi \propto \sigma_{\rm tot}$ with an IMF slope
of $x = 1.35$, with the four different infall prescriptions.}

\item
\label{fig:fets}
{Iron abundance as a function of time for different star formation
rates.  Data are from Edvardsson et al.\ (\protect\cite{edetal}).}

\item
\label{fig:imfs}
{For three different IMFs and SFR 5, evolution of (a) D, (b) \he3 and
D+\he3, and (c) \he4.}

\item
\label{fig:dheinf}
{D and \he3 evolution for the infall prescriptions.  Note the
enhancement of D and dilution of \he3 in the infall models.}

\item
\label{fig:yno}
{Low metallicities abundances of \he4, N, and O.  data is for
extragalactic \hii\ regions, as reported by Pagel et al.\
(\protect\cite{pageletal}) and Skillman et al.\ (\cite{skiletal}).
(a) N versus O.  (b) The \he4 mass fraction as a function of O/H at
low metallicities.  (c) \he4 as a function of N/H for different
treatments of N evolution.}

\item
\label{fig:run1}
{(a) Plot of the yields for the full set of 1184 candidate global
models.  Calculated for $\eta = 3 \times 10^{-10}$ and for a
particular set of light element features.  Abundances are calculated
for the birth of the solar system and compared to observed solar
abundances.  Solar abundances are from Geiss (\protect\cite{geiss})
for D and \he3, and from Anders \& Grevesse (\protect\cite{ag}) for
all other elements; $\pm 0.4$ dex limits to the solar abundances are
indicated by the dashed lines. \\ (b) Yields for the 267 candidate
models having CNOFe within 0.4 dex of solar abundances.}

\item
\label{fig:gdex}
{The G-dwarf distribution in two candidate global models.  The model
in (a) passes, the one in (b) does not; see discussion in text.  Data
is from Rana (\protect\cite{rana}).}

\item
\label{fig:run2-nocut}
{Solar abundances for the 2052 allowed global models with all light
element model variations.  No cut has been made for CNO overabundances
due to mass loss in massive stars (see discussion in text).}

\item
\label{fig:run2-cut}
{(a) As in figure \protect\ref{fig:run2-nocut}, with the constraint
that CNOFe be within 0.4 dex of solar; 1460 models displayed.  Note
that imposing this constraint makes little change in the light element
ranges. \\ (b) As in figure \protect\ref{fig:run2-cut}, for models
with $\eta = 4 \times 10^{-10}$.}

\item
\label{fig:run2-bigcut}
{The data of figure \protect\ref{fig:run2-cut} which reproduces CNOFe
within 0.3 dex (short dashed line).  For comparison, a level of 0.4
dex is shown in a long dashed line.}

\item
\label{fig:dshe3s}
{Plot of D and \he3 predicted solar abundances for the ``final set''
of 1460 models.}

\item
\label{fig:dshe4s}
{(a) (Pre)solar D versus \he4 for models in having no \he4 enhacement
features. \\ (b) Solar D and \he4 for models having either or both
\he4 enhacement features.}

\item
\label{fig:dihe3i}
{Calculated abundances of D and \he3 at the present epoch.  As
discussed in the text, it is unclear how to use the ISM \he3 data as a
constraint; we have omitted it here.  We show the D abundance in the
ISM due to Linsky et al.\ (\protect\cite{linsky}).}

\item
\label{fig:dshe3sdcut}
{Calculated D and \he3 abundances at the solar birth.  As in figure
\protect\ref{fig:dshe3s}, but models have been constrained to fit both
the presolar and the ISM D abundances.}

\item
\label{fig:dfez3o0}
{D versus [Fe/H] for our models, plotted at a galactic age equal to
universal time at $z=3$.  The redshift-time relation is derived for a
cosmology with $\Omega=0$ and $H_0 = 75$ km/s/Mpc.  Dotted lines
indicate primordial levels for the three adopted $\eta$ values.}

\item
\label{fig:dfeimf}
{D versus [Fe/H] for the models of figure \protect\ref{fig:imfs}
having different IMFs and for the high infall model 4 of table
\protect\ref{tab:gmodf}.  Regardless of these model features, the D
evolution at low metallicity is minimal.}

\item
\label{fig:ycnoop}
{Mass fraction of \he4 plotted as a function of (a) C, (b) N, and (c)
O abundance in the low metallicity regime.  Solid line: no \he4
enhancement; dashed line: $8-12$ \msol\ \he4 enhancement; dot-dashed
line: mass loss; dotted line, both enhancements.  For $\eta =
4\times10^{-10}$.}

\item
\label{fig:yerror}
{The error $\delta Y_{\rm p}$ made in extrapolating the $Y-$N relation
linearly to determine the primordial \he4 abundance.  Plotted for each
model as a function of the average slope $dY/d$N of the $Y-$N relation
at low metallicity.  Open circles are for models with only primary N,
triangles are models with mixed primary and secondary N, squares are
models with low secondary N and hexagons are for models with high
secondary N.  Horizontal line indicates level of observational error;
vertical line indicates approximate level of minimum slope for \hii\
region data.}

\end{enumerate}

\end{center}

\end{document}